\documentclass[conference]{IEEEtran}
\IEEEoverridecommandlockouts
\usepackage{cite}
\usepackage{amsmath,amssymb,amsfonts}
\usepackage{algorithmic}
\usepackage{graphicx}
\usepackage{textcomp}
\usepackage{xcolor}
\usepackage[caption=false,font=footnotesize]{subfig}
\usepackage{multirow}
\usepackage{array}
\usepackage{bm}
\usepackage{textgreek}
\usepackage{listings}
\usepackage[hidelinks]{hyperref}

\definecolor{keycolor}{rgb}{0,0.4,0.6}
\definecolor{stringcolor}{rgb}{0.6,0.1,0.1}
\definecolor{commentcolor}{rgb}{0.4,0.4,0.4}

\lstdefinelanguage{yaml}{
  keywords={true,false,null,y,n},
  keywordstyle=\color{darkgray}\bfseries,
  basicstyle=\ttfamily,
  sensitive=false,
  comment=[l]{\#},
  morecomment=[s]{/*}{*/},
  commentstyle=\color{purple}\ttfamily,
  stringstyle=\color{blue}\ttfamily,
  moredelim=[l][\color{orange}]{\&},
  moredelim=[l][\color{magenta}]{\*},
  moredelim=**[il][\color{red}{: \ }]{: \ },
  morestring=[b]',
  morestring=[b]"
}

\lstdefinestyle{hammer_yaml}{
    language=yaml,
    basicstyle=\ttfamily\scriptsize, 
    breaklines=true,
    frame=tb, 
    numbers=left,
    numberstyle=\tiny\color{gray},
    keywordstyle=\color{blue}\bfseries,
    stringstyle=\color{stringcolor},
    commentstyle=\color{commentcolor}\itshape,
    showstringspaces=false,
    tabsize=2,
    captionpos=b 
}

\def\BibTeX{{\rm B\kern-.05em{\sc i\kern-.025em b}\kern-.08em
    T\kern-.1667em\lower.7ex\hbox{E}\kern-.125emX}}
\begin{document}

\title{Testyard Benchmarks: a Dynamic RISC-V-based Benchmark Framework including DfT insertion, Structural and Functional Test Flows
\thanks{The \textit{Testyard} benchmarks repository can be found at: \url{https://github.com/cad-polito-it/testyard}. Corresponding authors: Francesco Angione, Nicola di Gruttola Giardino, Gabriele Filipponi, and Giusy Iaria.}
}

\author{\IEEEauthorblockN{Francesco Angione}
\IEEEauthorblockA{\textit{Dept. of Control and Computer Eng.} \\
\textit{Politecnico di Torino}\\
Turin, Italy \\
francesco.angione@polito.it}
\and
\IEEEauthorblockN{Nicola di Gruttola Giardino}
\IEEEauthorblockA{\textit{Dept. of Control and Computer Eng.} \\
\textit{Politecnico di Torino}\\
Turin, Italy \\
nicola.digruttola@polito.it}
\and
\IEEEauthorblockN{Gabriele Filipponi}
\IEEEauthorblockA{\textit{Dept. of Control and Computer Eng.} \\
\textit{Politecnico di Torino}\\
Turin, Italy \\
gabriele.filipponi@polito.it}
\and
\IEEEauthorblockN{Giusy Iaria}
\IEEEauthorblockA{\textit{Dept. of Control and Computer Eng.} \\
\textit{Politecnico di Torino}\\
Turin, Italy \\
giusy.iaria@polito.it}
\and
\IEEEauthorblockN{Irith Pomeranz}
\IEEEauthorblockA{\textit{School of Electrical and Computer Eng.} \\
\textit{Purdue University}\\
West Lafayette, IN, USA \\
pomeranz@purdue.edu}
\and
\IEEEauthorblockN{Paolo Bernardi}
\IEEEauthorblockA{\textit{Dept. of Control and Computer Eng.} \\
\textit{Politecnico di Torino}\\
Turin, Italy \\
paolo.bernardi@polito.it}
}

\maketitle

\begin{abstract}
The main challenge of modern computing manufacturing is to produce silicon products capable of operating under the worst possible conditions, both in data centers and under stringent functional safety constraints. These requirements impose exceptionally high quality standards on silicon vendors, while testing costs keep escalating as transistors continue to shrink. Evaluating new testing strategies, however, presents a dual challenge for researchers. Standard academic benchmarks, such as the ITC-99 suite, remain valuable for fast test prototyping, but they lack the structural complexity of modern VLSI devices. Assessing the true computational effort of academic methodologies on industrial scale designs is equally difficult without accessible and scalable models. A further obstacle is the absence of a unified framework able to abstract the commands imposed by each EDA tool and the configurations tied to each technology library, forcing researchers to spend time adapting and debugging fragile scripts. This work addresses both needs by proposing the \textit{Testyard} framework, built on top of the Chipyard framework, to derive a set of open-source, industrial scale benchmarks. Testyard also extends the Hammer VLSI flow abstraction framework with hardware testing capabilities. Three new functionalities are introduced into Hammer: Design-for-Testability (DfT) insertion, Automatic Test Pattern Generation (ATPG), and functional tests fault simulation. Each functionality is driven by high-level descriptions, while the framework translates these inputs into the commands required by each EDA tool, so that researchers no longer depend on a specific tool or technology library. The extended framework is applied to 13 representative SoC configurations from the Chipyard ecosystem, reporting the first comprehensive structural and functional test results across three technology libraries (45~nm, 15~nm, and 7~nm) and multiple fault models, including stuck-at, transition delay, and small delay faults. The resulting netlists, already provided with the DfT architecture, together with the extended Hammer flows, are publicly released on GitHub, offering the community a reproducible, open-source infrastructure for prototyping hardware testing methodologies at industrial scale complexity.
\end{abstract}


\section{Introduction}
Benchmarks are standardized models that enable rigorous and fair comparisons of algorithms, tools, and testing strategies within academia. Consequently, over the years, many suites of benchmark circuits have been proposed, such as ISCAS-85~\cite{ISCAS85}, ISCAS-89~\cite{ISCAS89}, ITC-99~\cite{ITC99}, and IWLS-05~\cite{IWLS2005}, to enable researchers to work on complex, open-source circuits while simultaneously providing valuable methodologies for the industry.

As the complexity of modern silicon devices increases exponentially, the evaluation of new testing strategies requires benchmarks that closely mirror industrial designs, enabling direct assessment of computational effort and methodology scalability on realistic circuits. At the same time, benchmarks must also remain manageable for fully
open-source academic tools since some features that may be needed to support a research direction are only accessible when the source code is available. Both requirements serve different research models: large, complex benchmarks ensure
that results are predictive of actual silicon behavior, while
smaller, manageable configurations support research that requires access to the internal functions or data structures of a testing tool. To support both models, there is a critical need for flexible, open-source benchmark suites capable of scaling seamlessly from smaller, manageable configurations up to high-complexity, industry-standard architectures~\cite{ANGIONE_BENCHMARKS_ETS_2026}.

Moreover, a crucial aspect when developing hardware testing methodologies, algorithms, and tools is the absence of a unified framework capable of abstracting the commands imposed by each EDA tool and the configurations tied to each technology library. Test engineers often need to manually adapt scripts for synthesis, Design-for-Testability (DfT) insertion, and static timing analysis from other workspaces or third-party repositories. Manually adapting these scripts to new designs is highly error-prone, yet the core methodologies often remain the same; the same applies when porting a methodology to a new technology library. Consequently, a majority of development time is devoted to adapting and debugging these scripts. This negatively affects productivity and can introduce hidden, hard-to-spot bugs.

\begin{figure}[!htbp]
    \centering
    \includegraphics[width=0.99\columnwidth]{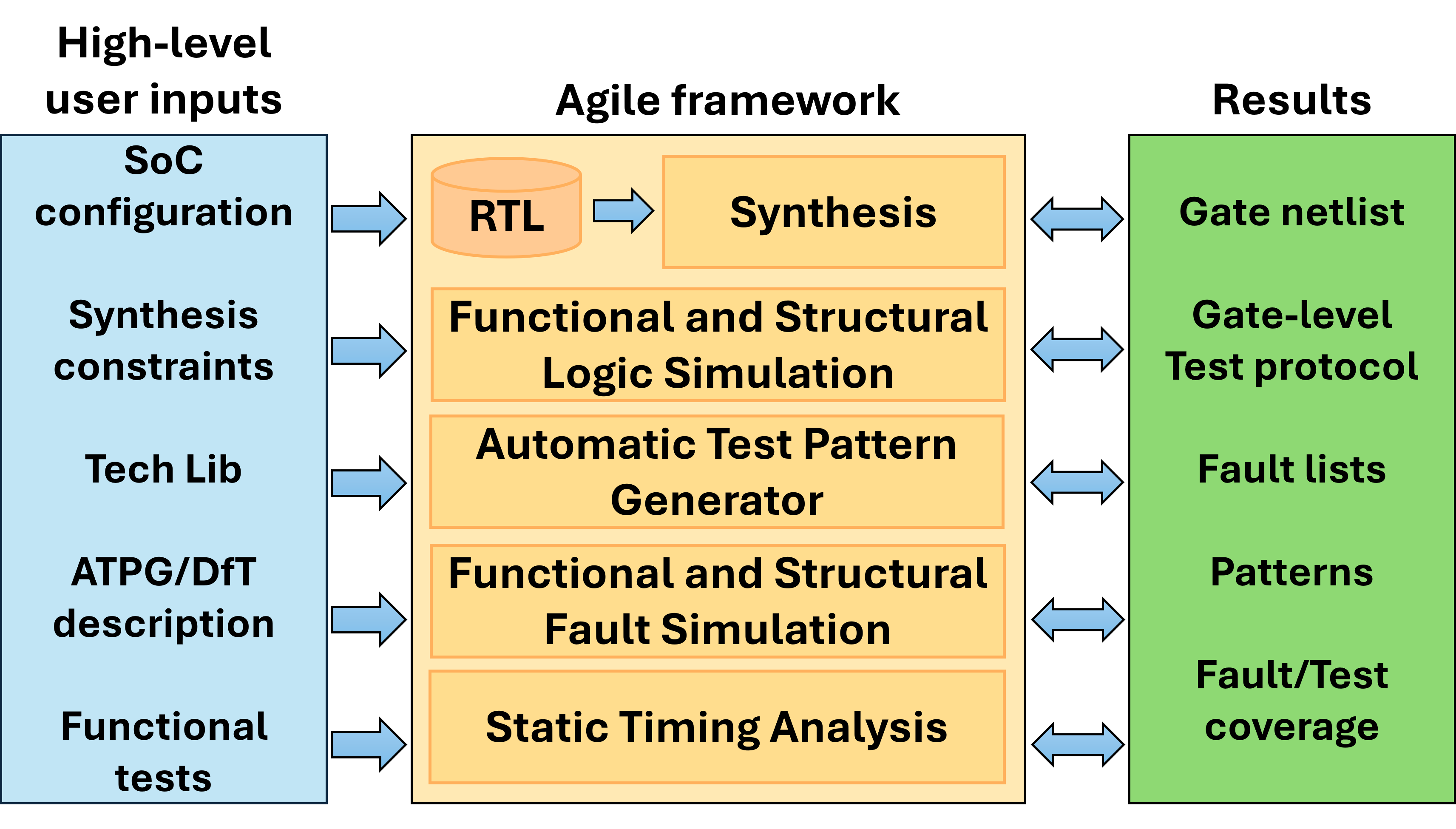}
    \caption{A conceptualized agile framework for prototyping hardware testing methodologies.}
    \label{fig:agile_framework}
\end{figure}

The burden of manual script adaptation can be overcome by introducing an agile framework, such as the one presented in Figure~\ref{fig:agile_framework}.

The conceptualized agile framework proposed in Figure~\ref{fig:agile_framework} aims to abstract EDA-dependent scripts for different phases of the VLSI flow (synthesis, DfT insertion, ATPG, etc.). The framework requires high-level user inputs—such as the target clock frequency, the number of scan chains, and the target technology library, in a user-friendly format. Internally, the framework handles all user inputs, executes the specific VLSI steps through internal communication between different tools and software modules, and provides the final reports. In this way, test researchers can focus on prototyping the methodology rather than "bug-hunting" in adapted scripts.

Therefore, two critical needs emerge:
\begin{itemize}
\item Modern benchmark suites comprising both complex designs similar to industry-strength circuits and more manageable designs suitable for academic tools.
\item An agile framework for prototyping hardware testing methodologies that abstracts the manual work of adapting fragile scripts between different tools or technology libraries and has a unified framework for functional and structural testing.
\end{itemize}

In this article, the aforementioned needs are addressed by proposing the \textit{Testyard} framework, that is based on the Chipyard~\cite{CHIPYARD} and Hammer~\cite{HAMMER} frameworks.

Chipyard~\cite{CHIPYARD} is an integrated framework for designing and evaluating complex Systems-on-Chip (SoCs) for computer architecture research purposes. It links open-source hardware generators, tools, and flows. The designs available in Chipyard are based on Chisel (a Scala-based hardware construction language) and use the Rocket Chip generator~\cite{ROCKETCHIP_GENERATOR}.
A library of ready to use IP is available, including:
\begin{itemize}
    \item In-order and out-of-order processors with different complexity.
    \item Accelerators for Neural Networks, vector processing and digital signal processing (DSP).
    \item Different types of interconnects.
\end{itemize}
The Chipyard framework tackles the flow from system to RTL design; meanwhile, the Hammer framework is in charge of the VLSI flow. Testyard uses Chipyard as a source for test benchmarks and extends Hammer with new hardware testing capabilities.

Hammer (Highly Agile Masks Made Effortlessly from RTL)~\cite{HAMMER} is a framework that abstracts the tedious parts of the VLSI flow (from RTL to physical design), such as synthesis, place and route, static timing analysis etc.
The framework abstracts the EDA tools by using Python-based wrappers; by leveraging high-level constraints, it can generate \textit{TCL} scripts (EDA tool dependent).
Furthermore, it provides abstraction of the technology library, allowing the same flow to be applied across various process nodes without scripts duplications.

This work significantly extends the Hammer framework to bridge the gap between physical design and hardware test. The contributions of this article include:
\begin{itemize}
    \item Implementation of the DfT insertion during synthesis.
    \item Implementation of the ATPG flow.
    \item Introduction of a functional fault simulation infrastructure to evaluate the efficacy of firmware-based tests.
\end{itemize}

By integrating these testing phases directly into the agile VLSI flow, this work enables researchers to evaluate different testing methodologies.

This paper presents the first comprehensive structural and functional test results on third-party designs, from the Chipyard ecosystem, ranging from small SoCs to complex Network-on-Chip (NoC) in order to provide a full picture of the testing scenario of functional and structural tests, for different fault models, such as stuck-at faults, transition delay, and small delay faults.
Experimental results provide insight into extensive ATPG generation and functional fault simulation campaigns for three technology libraries (45 $nm$, 15 $nm$ and 7 $nm$).

Furthermore, a key contribution of this work is the public release of the resulting synthesized netlists~\cite{CHIPYARD_CAD} and the hardware testing capabilities extension to the framework~\cite{HAMMER_CAD}. By providing the community with ready-to-use, DfT-inserted netlists for complex Chipyard-based designs, this paper establishes a new set of benchmarks that supports both academic evaluation and industry scale validation.

The remainder of the paper is organized as follows. Section~\ref{sec:background} presents background on the manufacturing test flow, the related structural and functional tests, the role of VLSI benchmarks in addressing both academic and industrial evaluation requirements, and an overview of the Chipyard and Hammer frameworks. Section~\ref{sec:testyard} presents the proposed \textit{Testyard} benchmark suite and details the DfT insertion, ATPG, and functional fault simulation actions added to the Hammer VLSI flow. Section~\ref{sec:experimental_results} presents the first comprehensive structural and functional testing results for these designs, evaluated across multiple technology nodes. Finally, Section~\ref{sec:conclusions} summarizes the findings and discusses future research directions.

\section{Background}
\label{sec:background}

This section provides the foundational context for hardware testing, beginning with an overview of the manufacturing test flow before detailing the primary structural and functional testing methodologies. Next, it outlines the role of VLSI benchmarks as essential frameworks for evaluating these testing strategies across both academic and industrial scales. Finally, it reviews the Chipyard and Hammer frameworks.

\subsection{Manufacturing test flow}

As the transistors scale, the density of transistors per area increases~\cite{CAMPBELL_COMPLEXITYANDTESTING_ETS_2010}. This important achievement led to an exponential rise in the complexity of integrated circuits, and the testing scenario becomes more complex~\cite{CAMPBELL_COMPLEXITYANDTESTING_ETS_2010}.

As previously mentioned, structural (scan-based) tests have been introduced to reduce the testing complexity and increase the automation of generating test patterns for a given integrated circuit~\cite{IEEE_BSANTAP_IEEE_2012}.
From the introduction of scan-based structural approaches, the generic manufacturing test flow has been stabilized, with minor changes in the order of test steps, which is company-dependent.

\begin{figure}[!htbp]
    \centering
    \includegraphics[width=1\columnwidth]{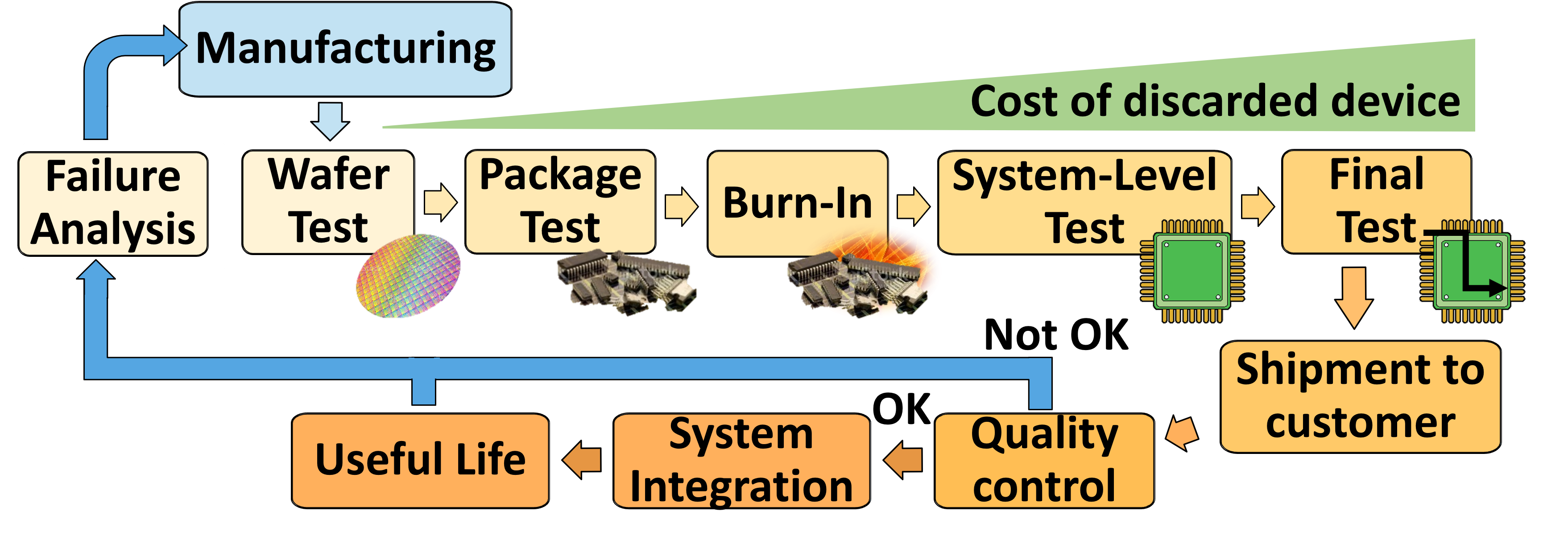}
    \caption{A possible manufacturing test flow.}
    \label{fig:manufacturing_test_flow}
\end{figure}
The test flow is divided into different stages as illustrated in Figure~\ref{fig:manufacturing_test_flow}, each one of them targeting different defects and with a specific goal to discard manufactured faulty devices~\cite{CHEN_WAVERLEVELSTRESS_ITC_2020,POLIAN_MYSTERIESSLT_ATS_2020}:
\begin{itemize}
    \item Wafer test done at the wafer level checks for the primary electrical functionalities of the chip and executes structural test patterns.

    \item Package test done at the package level measures and tests the essential electrical characteristics of the pins.
    \item Burn-in, mainly for automotive and safety-critical devices, has the goal of exacerbating latent defects~\cite{BENSO_ATPGBURNINFULLSCANCIRCUITS_ATS_2006,ANGIONE_BITOOLCHAIN_ACCESS_2023}. In this way, after the Burn-In phase, devices manufactured with weaknesses can be captured by later test steps.
    \item System-Level Test verifies the correctness of devices by resorting to complex functional programs~\cite{CHEN_RISINGNEEDSLT_VLSI_2018}.
    \item Final Test can detect faults by applying a mix of structural and functional tests.
\end{itemize}

As previously mentioned, the manufacturing test flow may result in test escapes since achieving the complete fault coverage is increasingly more complex. Furthermore, following the increasing complexity of modern SoCs, faults may also arise in the field due to different causes such as aging and marginalities. To monitor these risks, structural and functional tests are deployed in the field during the device's lifecycle. Failure to detect such field faults could cause a field-return scenario where manufacturing tests can be re-executed from the manufacturer, possibly causing a No-Trouble-Found (NTF)~\cite{POLIAN_MYSTERIESSLT_ATS_2020}, or they can be labeled as Silent Data Corruptions~\cite{DIXIT_SDCWILD_ARXIV_2022,HOCHSCHILD_CORESDONTCOUNT_HOTOS_2021,ANGIONE_SDCTESTESCAPES_ITCINDIA_2025}.

The manufacturing test phases discussed above rely on two distinct paradigms: structural testing, which targets physical hardware defects, and functional testing, which verifies system-level behavior. A further discussion of these two core methodologies is presented in the following sections.

\subsection{Structural Testing}
Structural tests are mainly executed during the manufacturing test flow, and to some extent, during Power-On Self-Test in the field~\cite{BERNARDI_LOGICDIAGNOSIS_MDPI_2024, BERNARDI_ENHANCINGLOGICDIAGNOSIS_LATS_2025}. They mainly reconfigure the device in a non-functional fashion to achieve the highest coverage of hardware defects. Additionally, structural testing can also be performed during runtime in the field using dedicated technologies such as Intel In-Field Scan (IFS)~\cite{INTEL_IFS}, which allows for periodic health checks without disrupting the system's main operations.

\begin{figure}[!htbp]
    \centering
    \includegraphics[width=\columnwidth]{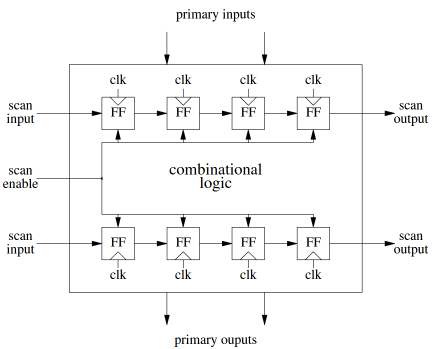}
    \caption{Sequential circuit example for scan-based test generation.}
    \label{fig:atpg_circuit}
\end{figure}

Fundamentally, structural testing relies on scan architectures that interconnect the internal flip-flops into shift registers, thereby enabling direct control and observation of the internal states. Because of these capabilities, scan-based testing is a crucial approach for achieving high fault coverage in digital designs. The Automatic Test Pattern Generation (ATPG) is a powerful solution for test pattern generation oriented to test digital circuits.
Figure~\ref{fig:atpg_circuit} illustrates a sequential circuit, including both combinational logic and flip-flops, along with the flow used for testing manufactured devices.

As shown in Figure~\ref{fig:atpg_circuit}, scan-based patterns are based on loading precomputed values into the scan chains, executing those values at speed through the combinational logic, and subsequently capturing the computed values. These values are then shifted out for comparison with the golden computed values generated by the ATPG, while new patterns are shifted in.

To meet the demands for reliability and adherence to stringent requirements dictated by quality standards, designers integrate additional circuitry capable of conducting on-device self-tests, formally known as Built-In Self-Test (BIST)~\cite{10.5555/34513, Abramovici1990DigitalST, Bardell1982SelfTestingOM}. When the self-test targets the combinational part of the device, the BIST is called Logic BIST (LBIST)~\cite{10.5555/34513, Abramovici1990DigitalST}. In the standard self-testing using a Multi Input Shift Register (MISR) and parallel Shift Register Sequence Generator (SRSG) (STUMPS) architecture~\cite{10.5555/34513, Abramovici1990DigitalST, Bardell1982SelfTestingOM}, a Linear Feedback Shift Register (LFSR)~\cite{10.5555/34513} is the engine for input pattern generation, whereas a signature representing the test outcome is produced through a MISR~\cite{10.5555/34513}. To identify irregularities, the final signature accumulated at the end of the test flow is compared with a known reference called the golden signature.

Ultimately, structural tests remain the indispensable foundation for achieving high fault coverage and driving production quality. When necessary, functional tests can be implemented in subsequent phases as an additional layer to verify complex system-level interactions. This synergy~\cite{IARIA_MULTIPHASEDIAGNOSIS_ACCESS_2026, IARIA_SEU_ISIE_2022} ensures maximum reliability for modern high-density digital designs.

\subsection{Functional Testing}

Structural methods primarily focus on the integrity of individual gates and paths, yet they can fail to replicate the functional interactions typical of real-world workloads. Furthermore, structural testing may not be entirely exhaustive for modern, high-density designs, particularly regarding complex timing-oriented faults, which necessitates the use of complementary testing strategies to ensure full coverage~\cite{CHEN_RISINGNEEDSLT_VLSI_2018}. Consequently, functional testing is integrated into the later stages of the manufacturing flow~\cite{ANGIONE_SLTXBAR_TOC_2024} and deployed as online testing strategies dictated by functional safety standards. Under the umbrella of functional testing, various methodologies exist across different stages of the device lifecycle, primarily executing firmware-oriented hardware tests.

One methodology is Software-Based Self-Test (SBST), which runs functional test patterns based on the processor instruction set, exploiting processor resources to test the processor itself and its surrounding components~\cite{PASCHALIS_SBSTPERIODICPROCESSOR_TCAD_2005,PARVATHALA_FRITIS_USPATENT_2015,PIUMATTI_DEVELOPEMENTSBSTMCUFAMILY_MR_2020}. Widely adopted within industrial manufacturing flows~\cite{BAYRAKTAROGLU_CACHEFUNCTIONALMICROPROCESSORTESTING_ITC_2006}, SBST is also an emerging alternative for performing online functional testing during the in-field application of microprocessor-based systems.

Functional testing principles are also applied during the Burn-In (BI) phase, which is a stress phase designed to screen out infant mortality defects in manufactured devices~\cite{MAK_INFANTMORTALITYRELIABILITYISSUE_IOLTS_2007}. This phase combines external environmental stressors, such as thermal acceleration~\cite{ZAKARIA_REDUCINGBURNINTIMEHIGHVOLTAGESTRESS_DTC_2006}, with internal electrical stress to drive circuit nodes into high-toggle activity modes~\cite{BENSO_ATPGBURNINFULLSCANCIRCUITS_ATS_2006,APPELLO_OPTIMIZEDBURNINAUTOMOTIVESOC_DT_2018}. Creating functional stress routines that produce appropriate internal activity is often less manually intensive than writing directed SBST programs, as the generation process can be automated using randomized tools or evolutionary engines to yield high-quality stress patterns with minimal human effort~\cite{BERNARDI_IMPROVINGSTRESSQUALITY_DVRBOOK_2017}.

In the last decade, System-Level Test (SLT) has been widely integrated as a mandatory manufacturing step to run functional applications that mimic the exact in-field behavior of the device under test~\cite{APPELLO_SLTSTATEOFTHEART_IOLTS_2021,TIPPARTHI_CONCURRENTSLT_EPTC_2014,BISWAS_INDUSTRIALSLT_DTC_2012}. A common baseline benchmark involves booting and executing a Real-Time Operating System (RTOS) under typical operational workloads. Modern automation techniques have emerged to synthesize these complex workloads, including grey-box program generation methods that leverage fuzzing based on mutation to control processor power consumption constraints~\cite{SCHWACHHOFER_GREYBOX_ETS_2023,SCHWACHHOFER_UGP4SLT_ETS_2024}. Other approaches utilize Large Language Models (LLMs) with Structural Chain of Thought (SCoT) prompting to generate high-quality SLT code targeting specific non-functional properties like Instructions Per Clock (IPC)~\cite{SCHAWCHHOFER_LLMSLT_DFT_2024}. Crucially, SLT environments verify communication peripherals and un-core logic by evaluating complete system-level interactions between the CPU, memories, and external peripherals. By providing this holistic validation environment, automated SLT generation methodologies can overcome the structural test weaknesses and lack of peripheral interaction observed in traditional manufacturing flows~\cite{ANGIONE_SLTCAN_TOC_2025,ANGIONE_SLTXBAR_TOC_2024}.

\subsection{The importance of VLSI Benchmarks for Hardware Testing}

During decades of academic research in the area of VLSI testing, benchmark circuits have been used for demonstrating the effects of new advancements in both software and hardware. These advancements include new algorithms for tasks such as fault simulation, test generation and test compaction, algorithmic improvements to existing approaches, new design-for-testability and synthesis-for-testability features, and new built-in self-test approaches.

Ideally, state-of-the-art designs and software tools would be open-source, allowing VLSI testing research to be performed in the most advanced software environment. However, the most advanced software environment would also require significant software engineering effort that is typically not manageable for academic research. The status quo that emerged supports two models of academic research.

Model 1 uses the available executable code of commercial software tools that can handle industry-strength designs. It uses features that the tools make available as output to demonstrate new research results. The work presented in this article focuses on closing the gaps that exist in this environment, attempting to ensure that both the software environment and the benchmarks are close to the industrial state-of-the-art.

Model 1 becomes cumbersome and even infeasible when the research requires access to internal data structures or functions of the software tool. For example, if an algorithm requires knowledge of the set of faults detected by each test, the information can easily be accessed in the internal data structures of a fault simulation or test generation tool; the executable code would have to be run multiple times to obtain this information. As another example, if the goal is to modify an internal function of a structural deterministic (branch-and-bound) test generation tool, such as the line justification function, this can only be performed using the source code.

To avoid limiting academic research to model 1, model 2 uses available academic software tools with full access to the source code. The academic software tools do not have the sophistication of the commercial tools, and cannot handle industry-strength designs, though they offer full algorithmic transparency and remain highly effective for a wide class of benchmarks. Nevertheless, over the years, innovations that were demonstrated using academic tools and benchmark circuits from the ISCAS-89, ITC-99 and IWLS-05 suites have influenced industrial practices.

Whereas the rest of this article focuses on closing the gaps for model 1, the proposed benchmarks include circuits of manageable sizes as well, ensuring compatibility with academic tools and thus supporting both models. The intention is not to eliminate model 2 that has resulted in many innovations over the years, but to offer the option to use model 1 with state-of-the-art designs.

\subsection{The Chipyard framework}
\label{sec:benchmarks}

The Chipyard framework~\cite{CHIPYARD} is a unified platform for architecture research and SoC development following an agile-oriented approach. It describes various hardware components, such as cores, accelerators, and memory-mapped peripherals, using the Chisel hardware description language embedded in Scala~\cite{CHISEL}.

\begin{figure}[!htbp]
    \centering
    \includegraphics[width=1\columnwidth]{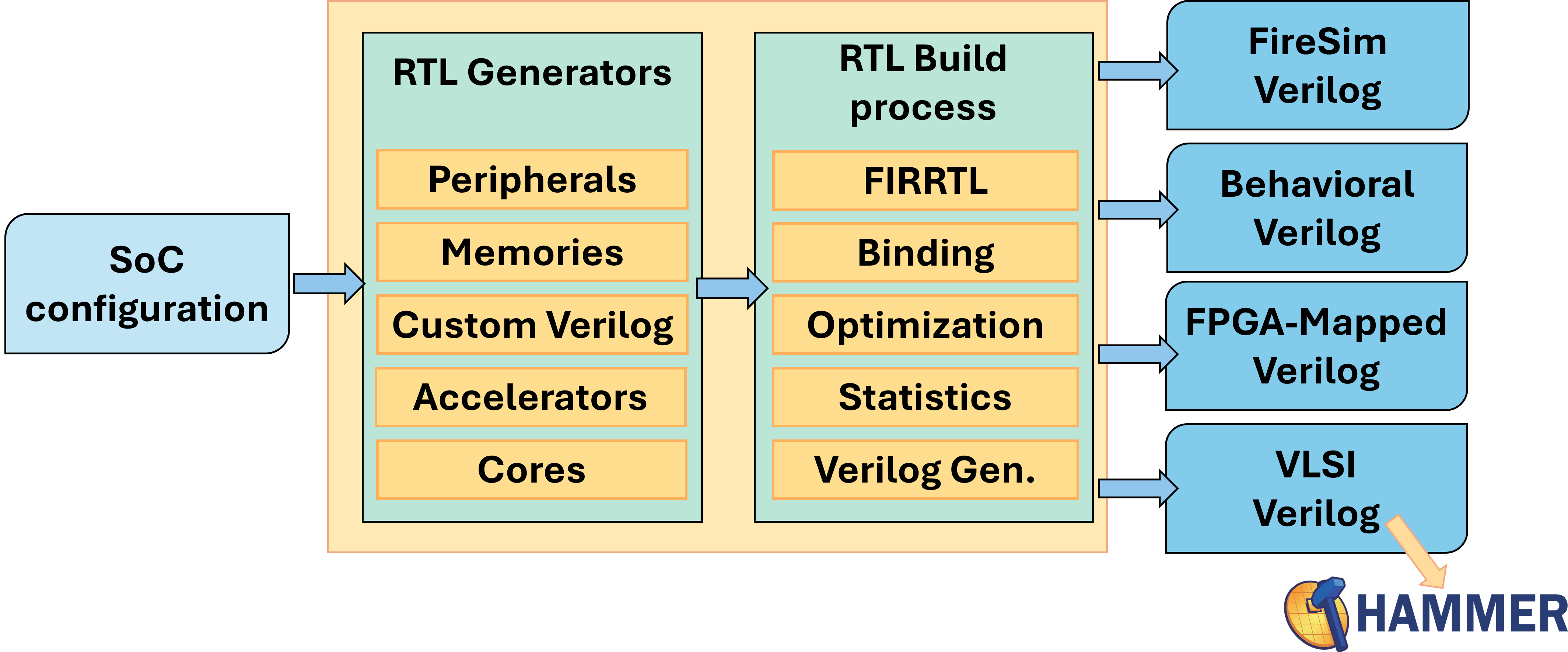}
    \caption{A simplified flow of the Chipyard framework (in yellow) and its outputs (in blue).}
    \label{fig:chipyard_flow}
\end{figure}

Figure~\ref{fig:chipyard_flow} represents a simplified view of the Chipyard framework.

Starting from an SoC configuration described in \textit{Scala}, the framework generates an intermediate circuit representation in \textit{CIRCT} (Circuit Intermediate Representation Compiler and Tools~\cite{CIRCT}), which incorporates \textit{FIRRTL} (Flexible Intermediate Representation for RTL~\cite{FIRRTL, FIRRTL2}). The intermediate representation is used to optimize the design, perform binding, and extract statistics of the SoC configuration. Afterward, Verilog files for different options are generated.

The Chipyard framework generates Verilog files for various purposes: FireSim for accelerated FPGA emulation~\cite{FIRESIM}, FPGA-oriented versions for prototyping, behavioral Verilog for RTL logic simulation, and Verilog suitable for the VLSI flow. The latter is processed by the Hammer framework to transform the RTL description into a GDSII layout~\cite{HAMMER}.

The strength of the Chipyard framework lies in its unified ecosystem, which integrates diverse hardware components and software simulation flows. These capabilities span from simple bare-metal applications to complex Real-Time Operating Systems (RTOSs) and Linux-based OSs, all ready for simulation on FPGAs, logic simulators, or instruction-set simulators.

Consequently, the integration of the Chipyard frameworks enables the exploration of how various architectural and core configurations impact a consistent VLSI flow, from a hardware testing perspective. This allows hardware testing and reliability researchers to investigate different methodologies within the same architecture or to evaluate the scalability and behavior of their approaches across diverse SoC architectures of different complexities.

\subsection{The Hammer framework}
\label{sec:hammer}

The Highly Agile Masks Made Effortlessly from RTL (Hammer) framework is a Python-based abstraction layer for the VLSI flow; it removes the burden of manually adapting every script, such as those for static timing analysis and synthesis, for each design and technology library. Hammer leverages high-level descriptions of timing and design constraints (in YAML, a human-readable data serialization language~\cite{YAML}), which it automatically translates into the specific commands required by EDA tools.

\begin{figure}[!htbp]
    \centering
    \includegraphics[width=1\columnwidth]{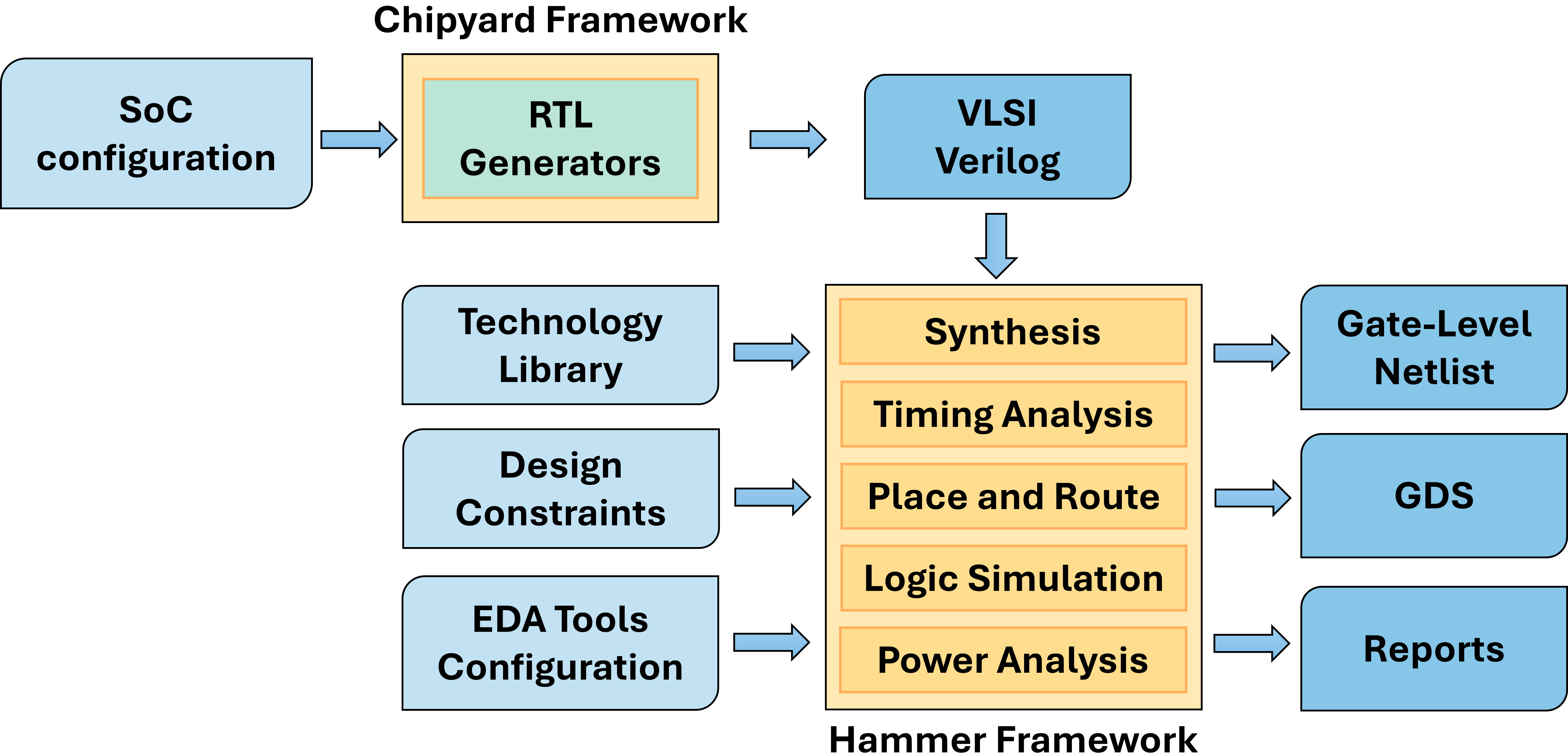}
    \caption{A simplified flow of the Chipyard-Hammer framework.}
    \label{fig:chipyard_hammer}
\end{figure}

Figure~\ref{fig:chipyard_hammer} illustrates the Chipyard-Hammer pipeline for the VLSI flow, starting from the generation of suitable Verilog for logic synthesis through to layout placement and routing. The Hammer workflow in Figure~\ref{fig:chipyard_hammer} demonstrates a simplified user input configuration for controlling these subsequent VLSI steps.

Design constraints, technology libraries, and EDA tool configurations are described in a high-level format to abstract specific EDA tool commands. The Hammer framework generates various outputs, ranging from synthesis, static timing analysis, and power analysis reports to the gate-level netlist of the generated VLSI Verilog and its physical implementation.

Although the Hammer framework is highly flexible, it currently lacks several fundamental VLSI flow steps for hardware testing.
These include DfT insertion during the synthesis process, ATPG, and the capability to perform functional fault simulation for functional test programs.

\begin{figure}
    \centering
    \includegraphics[width=0.6\columnwidth]{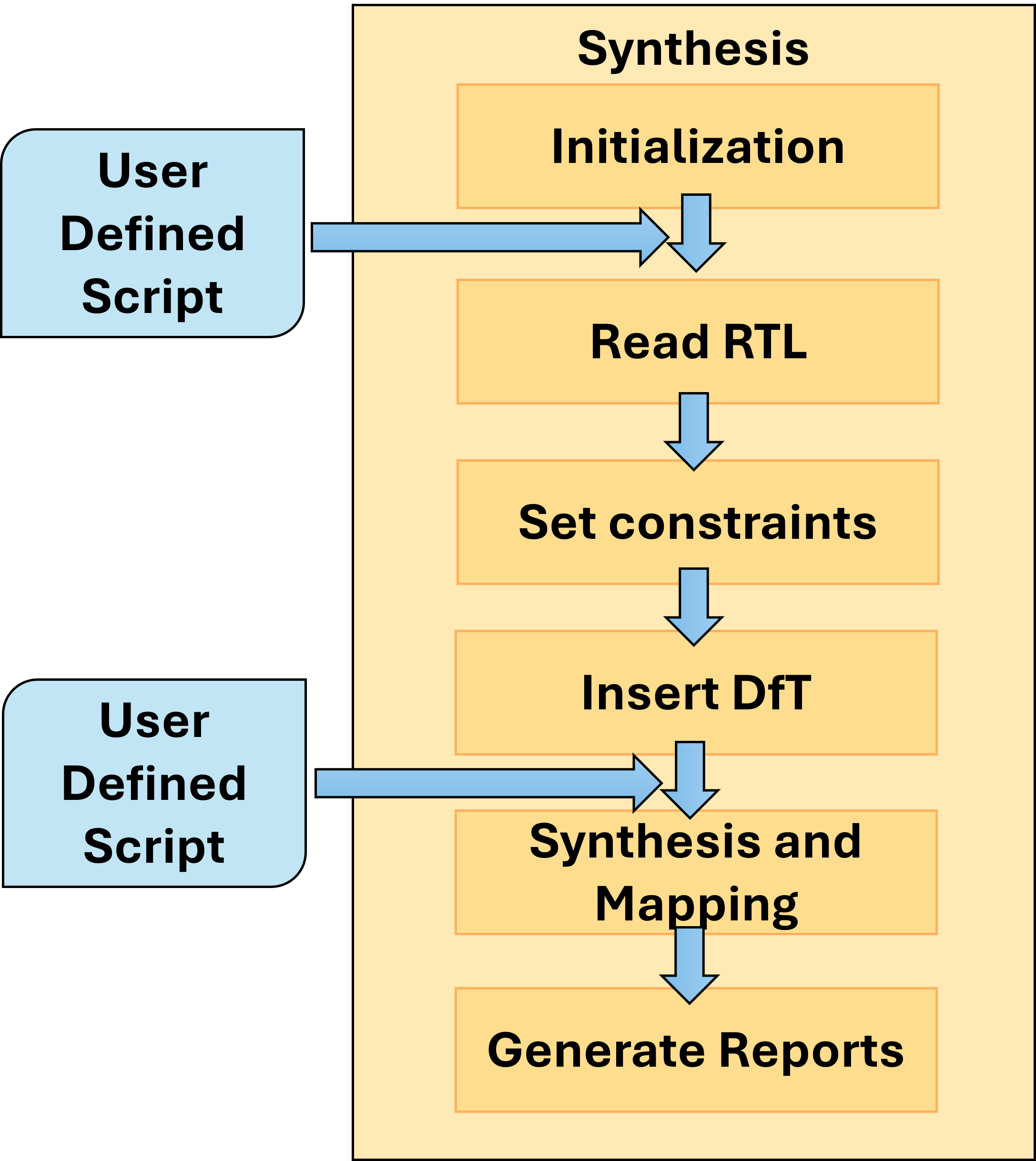}
    \caption{Detail of the Synthesis action in Hammer including User Defined scripts. }
    \label{fig:action_hook}
\end{figure}

To understand how these workflows manage execution internally, it is worth examining Hammer's underlying step structure. Any top-level Hammer action is natively decomposed into a sequence of discrete, addressable substeps. To accommodate custom design variations or user-defined scripts, the framework exposes a programmatic mechanism known as \textit{hooks}. As illustrated in Figure~\ref{fig:action_hook}, hooks act as targeted escape hatches within an execution block, such as the synthesis action, enabling developers to inject custom Python methods or user-defined scripts before, after, or in place of any default substep. Hooks expose the full flexibility of the framework for non-standard design requirements. For example, a user may inject a pre-synthesis hook to apply custom netlist patches, or a post-synthesis hook to invoke a proprietary, or customized, script for extracting data from the synthesized design. This mechanism ensures the framework remains extensible without requiring modifications to the core Hammer codebase.

\section{Testyard: Proposed benchmark suite and the related testing functionalities}
\label{sec:testyard}

\textit{Testyard} is an evolving suite of RISC-V-based SoC benchmarks for hardware testing research. Starting from a representative subset of Chipyard SoC configurations, each design is synthesized to a gate-level netlist with DfT inserted. These netlists, together with the corresponding DfT test protocol are publicly released. It is important to highlight that fault lists and structural or functional tests (except for the functional test programs already provided in the original repository) are not released, as we would like to avoid favoring one specific EDA vendor over another.

Beyond the benchmarks themselves, the flow used to generate them is provided. Indeed, this work extends the Hammer VLSI flow abstraction framework with three new actions (following Hammer's nomenclature for each step in the VLSI flow) dedicated to hardware testing: DfT insertion, Automatic Test Pattern Generation (ATPG), and functional fault simulation. These actions are exposed through the same high-level, YAML-driven configuration approach already used by Hammer for synthesis and place-and-route. This gives researchers the flexibility to go beyond the case studies proposed in this work: the same flow can be applied to other designs, other technology libraries, or other fault models.

Thus, the \textit{Testyard} framework, built on top of Chipyard and Hammer frameworks, aims at:
\begin{itemize}
    \item Proposing benchmark circuits of industrial scale, while still keeping smaller configurations available for compatibility with academic tools.
    \item Making hardware testing functionalities -- DfT insertion, ATPG, and functional fault simulation -- directly available within an existing EDA-tool abstraction framework, removing the need for ad hoc, tool and library-specific scripting.
\end{itemize}

\begin{table*}[ht]
    \centering
    \caption{Class and architectural description of the SoC configurations selected as Testyard benchmark case studies.}
    \label{tab:case_studies_architectural_description}
    \begin{tabular}{|l|l|l|l|l|}
        \hline
        \textbf{Class} & \textbf{Configuration Name} & \textbf{Core} & \textbf{Memories} & \textbf{HW Acc. \& Peripherals} \\
        \hline
        \hline
        Coprocessor & \texttt{AES256ECBRocketConfig} & Rocket & L1 64KB, L2 512KB & AES RoCC Accel, UART \\
        \hline
        CPU & \texttt{CVA6Config} & CVA6 (Ariane) & L1 16KB, L2 512KB & UART \\
        \hline
        CPU & \texttt{MediumBoomV3Config} & BOOMv3 (OoO) & L1 32KB, L2 512KB & UART \\
        \hline
        CPU & \texttt{MediumBoomV4Config} & BOOMv4 (OoO)& L1 32KB, L2 512KB & UART \\
        \hline
        HW accelerator & \texttt{GemminiRocketConfig} & Rocket & L1 64KB, L2 512KB & Gemmini Systolic Array, UART \\
        \hline
        CPU & \texttt{IbexConfig} & Ibex & L2 512KB & UART \\
        \hline
        Coprocessor & \texttt{FFTRocketConfig} & Rocket & L1 64KB, L2 512KB & FFT RoCC Accel., UART \\
        \hline
        CPU & \texttt{ShuttleConfig} & Shuttle & L1 48KB, L2 512KB & UART \\
        \hline
        Network on chip & \texttt{MultiNoCConfig} & Rocket & Banking L2 Cache & UART \\
        \hline
        CPU & \texttt{RocketConfig} & Rocket & L1 64KB, L2 512KB & UART \\
        \hline
        MCU & \texttt{ManyPeripheralsRocketConfig} & Rocket & L1 64KB, L2 512KB & GPIO, UART, SPI, I2C \\
        \hline
        Vector Unit & \texttt{REFV256D128M64RocketConfig} & Rocket & L1 64KB, L2 512KB, Vector Cache & Ref. Vector Accel., UART \\
        \hline
        CPU & \texttt{DualRocketConfig} & Rocket & L1 64KB (each core), L2 512 KB & UART \\
        \hline
    \end{tabular}
\end{table*}

\subsection{Proposed benchmark suite}
The available SoC designs in the Chipyard framework are used in the experimental section as case studies. It is crucial to highlight that the framework permits describing SoCs of arbitrary complexity, creating a perfect playground for evaluating the scalability of new proposed hardware testing methodologies.

Table~\ref{tab:case_studies_architectural_description} outlines the architectural specifications for various predefined SoC configurations within the Chipyard framework. In the Chipyard framework, there are more than 150 SoC configurations, with different architectural parameters and IP. A subset of 13 representative classes of SoCs has been selected, ranging from pure CPU-based SoC, Microcontrollers (MCU), Chiplet, CPU with coprocessor and vector unit, as well as network on chip, with and without hardware accelerator.

All of the configurations support debug over JTAG and RV64GC ISA, meanwhile communication across the SoC is achieved by means of a TileLink bus protocol.

The first column of Table~\ref{tab:case_studies_architectural_description} (`Class') identifies the class of the case study, while the second column (`Configuration') specifies the configuration name used in the Chipyard framework. The third column (`Core') shows the CPU used in the configuration.
The column `Memories' represents the amount of available memories in the SoC, including caches and RAMs. Meanwhile, column `HW Acc. \&  Peripherals' lists the available peripherals and hardware accelerator in the configuration.

The various classes go from simple CPU configurations (Rocket, Ibex cores) to more complex CPU configurations (BOOM cores), to classes with GP-GPU (Radiance GPU), to micro-controller configurations (MCU) with many peripherals or hardware accelerators to coprocessor and vector units.

Table~\ref{tab:riscv_cores} summarizes the SoC CPU configurations, from the simple five-stage in-order Rocket core to the high-performance out-of-order BOOM core.

\begin{table*}[htbp]
\centering
\caption{Comparison of RISC-V core microarchitectures within the Chipyard framework.}
\label{tab:riscv_cores}
\begin{tabular}{|c|c|c|c|p{4cm}|}
\hline
\textbf{Core} & \textbf{Execution Type} & \textbf{Issue Width} & \textbf{ISA Support} & \textbf{Typical Use Case} \\ \hline
\hline
Ibex       & In-order     & 1-wide       & RV32IMC & Embedded/IoT/Security \\ \hline
Rocket     & In-order     & 1-wide       & RV64GC  & General Purpose / Linux \\ \hline
Shuttle    & In-order     & 2-wide+      & RV64GC  & Superscalar Research \\ \hline
BOOM       & Out-of-Order & 1 to 5-wide  & RV64GC  & High-performance Research \\ \hline
\end{tabular}
\end{table*}

The Rocket core serves as the standard 64-bit (RV64GC) core, utilizing a 5-stage in-order pipeline with a non-blocking D-cache and an MMU to support complex operating systems like Linux. For designs requiring higher instruction throughput without the complexity of out-of-order execution, the Shuttle core provides a 7-stage superscalar in-order alternative.

The Ibex core~\cite{IBEXRISCV} (formerly Zero-riscy) provides a production-hardened 2 or 3-stage pipeline specifically optimized for IoT security and low-power control, as demonstrated by its deployment in the OpenTitan project~\cite{PARISI_OPENTITAN_ARXIV_2024}. While this core typically adopts the RV32 ISA to minimize area, it provides the necessary precision for sequencers and power management units within larger SoCs.

The Berkeley Out-of-Order Machine (BOOM)~\cite{ZHAO_BOOM_CARISCV_2020} represents the high-performance apex of the ecosystem, implementing a sophisticated superscalar, out-of-order execution engine for advanced computer architecture research. There are different micro-architectural configurations for the BOOM; however they mainly change the width of the various stages and size of the Reorder Buffer.

In an AI-oriented research context, the integration of deep learning accelerators within the Chipyard SoC framework is characterized by the management of data movement and computational density. The NVIDIA Deep Learning Accelerator~\cite{NVDLA} (NVDLA) configurations represent a scalable approach to fixed-function inference, primarily categorized into small and large variants. The small NVDLA configuration is optimized for area-constrained IoT applications, implementing a reduced Multiply-Accumulate (MAC) array and a simplified data path. A critical limitation of this variant is its reliance on a single memory interface connected to the system FrontBus, which forces weights and intermediate feature maps to compete for bandwidth with the host Rocket core's access to L2 cache and DRAM.

Conversely, the large NVDLA configuration scales computational throughput significantly—typically by a factor of 16—and incorporates specialized hardware engines for Winograd convolutions. To sustain this increased compute density, the architecture utilizes Dual Data Backbones. This structural enhancement provides a dedicated interface for a private local SRAM, effectively decoupling heavy weight-fetching and intermediate result storage from the primary system bus. This dual-interface strategy significantly increases effective bandwidth and prevents the accelerator from clogging the main system interconnect during high-throughput inference tasks. The small NVDLA configuration has been selected for the scope of this work.

The Gemmini accelerator~\cite{GEMMINI} offers a distinct architectural philosophy centered on a systolic array. Unlike the modular engine approach of NVDLA, Gemmini utilizes a decoupled access-execute architecture that integrates with the Rocket core via the RISC-V Custom Code (RoCC) interface. This tight integration allows for low-latency control through custom ISA extensions, whereas NVDLA is managed via memory-mapped I/O. Furthermore, Gemmini employs a hardware-managed scratchpad and an asynchronous DMA controller, providing researchers with the flexibility to explore various dataflow patterns, such as weight-stationary or output-stationary mappings. While NVDLA provides high efficiency for established convolutional neural network (CNN) structures, Gemmini serves as a more programmable research platform for evaluating novel tensor-based operations and fine-grained data orchestration within the RISC-V ecosystem. The standard Gemmini configuration has been selected, a 16$\times$16 mesh of Processing Elements (PEs), providing a balanced trade-off between computational density and hardware area. The configuration utilizes 8-bit signed integer (SInt) inputs and weights, while maintaining numerical precision through a 32-bit signed integer accumulator.

The execution model of the standard Gemmini configuration follows an output-stationary dataflow. Memory management is handled through a hardware-managed 256 KB scratchpad, which facilitates the staging of tensors for the systolic array. Furthermore, the accelerator is integrated into the SoC fabric via the TileLink protocol, allowing its internal DMA controller to communicate directly with the system bus. This integration enables asynchronous data transfers between main memory and the local scratchpad, ensuring the host Rocket core can manage the accelerator via the RoCC interface without saturating the processor's primary memory path.

Vector units in Chipyard, such as the Ara~\cite{ARA}, are designed to accelerate data-parallel workloads by implementing the RISC-V Vector (RVV) extension. Unlike standard scalar processors that operate on single data points, these units utilize a large vector register file to perform Single Instruction, Multiple Data (SIMD) operations.

The Coprocessor class identifies accelerators integrated via the Rocket Custom Co-processor (RoCC) interface. This interface provides a high-bandwidth, low-latency connection that allows specialized hardware—such as the AES256, FFT, or Zstd units—to be tightly coupled with the Rocket core's pipeline. By using the RoCC interface, these accelerators can intercept custom instructions directly from the instruction stream, sharing the host core's L1 data cache and virtual memory system. This architectural choice minimizes the software overhead associated with traditional memory-mapped I/O and allows the accelerator to operate on shared data structures with hardware-managed coherency.

The \textit{MultiNoCConfig} shifts the architectural focus from individual compute elements to the communication fabric that links them. In complex SoCs with high core counts and multiple accelerators, a simple crossbar becomes a bottleneck. The NoC implements a routed, packet-switched topology—such as a 2D mesh—to facilitate parallel communication between sources and destinations. As seen in the configuration table, these designs often utilize Banking L2 Caches, where the L2 capacity is distributed across the network. This distribution allows for multiple simultaneous memory requests and reduces hot-spots, ensuring that the interconnect can support the aggregate bandwidth demands of many-core environments and data-hungry hardware accelerators.

\subsection{Testing functionalities}

This subsection presents an extension to the Hammer framework incorporating missing and essential stages for hardware testing. In the following sections, each newly developed functionality integrated into the flow, named 'action' (following Hammer framework nomenclature for each step in the VLSI flow) is described in detail.

In each following described action, the presence of a target EDA tool dependent configuration is identified by the
\texttt{<tool>} placeholder in the YAML namespace;
in a real configuration, this is replaced by the name of the
specific EDA tool plugin selected for that action.
\subsubsection{Design-for-Testability Insertion}
Design-for-Testability insertion during the synthesis process is a fundamental step in hardware testing. This process introduces specific scan flip-flops that allow ATPG tools to effectively generate high-coverage test patterns. By providing full observability and controllability of internal flip-flops, it reduces the complexity of pattern generation at the cost of some area overhead that is widely accepted by the industry.

\begin{figure}[!htbp]
    \centering
    \includegraphics[width=0.8\columnwidth]{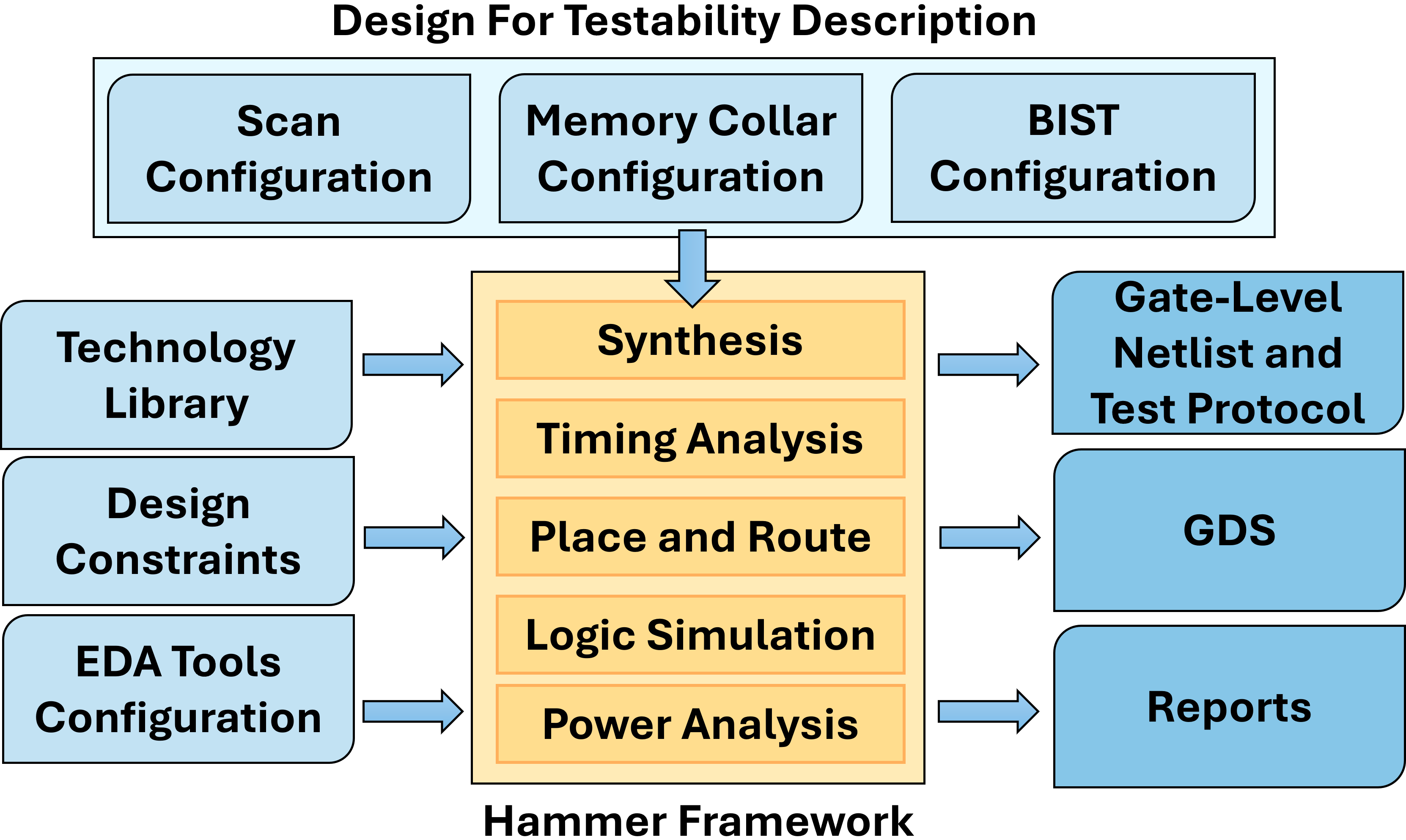}
    \caption{Design-for-Testability Insertion action (part of the synthesis action) in Hammer framework.}
    \label{fig:dft}
\end{figure}

Figure~\ref{fig:dft} presents the DfT description provided to the Hammer framework during the synthesis action.

This functionality in the synthesis action provides a high-level description of DfT architectures, specifically regarding scan insertion, memory collar integration, and BIST modules. A specific DfT component, such as a BIST or memory collar, can be added or removed simply by modifying this high-level description. Furthermore, the scan chain configuration, including the number of chains, flip-flop style, and control/observe signals, can be adjusted at this level. These descriptions are then translated into EDA-specific DfT insertion commands, which are executed during the synthesis step (producing a gate-level netlist with DfT architecture) as a prerequisite for the ATPG process.

\begin{figure}[htbp]
    \centering
    \begin{minipage}{0.9\columnwidth}
    \begin{lstlisting}[style=hammer_yaml]
# DfT insertion option
synthesis.dft_insertion : true 

# Scan configuration 
synthesis.insert_dft.scan : true 
synthesis.dft.scan.style : "multiplexed_flip_flop" 
synthesis.dft.scan.chain_count : 1
synthesis.dft.scan.ports : [
  {"name" : "jtag_TDI" , "direction" : "in" ,
   "type": "ScanDataIn" , "exist" : true },
  {"name" : "jtag_TDO" , "direction" : "out" ,
   "type": "ScanDataOut" , "exist" : true }
]
synthesis.dft.scan.clock_ports: [
  {"name" : "jtag_TCK", "active_state" : "1",
   "timings" : ["45" , "95"] },
  {"name" : "clock_uncore", "active_state" : "1",
   "timings" : ["45" , "95"] },
  {"name" : "serial_tl_0_clock_in", "active_state" : "1",
   "timings" : ["45" , "95"] }
]
synthesis.dft.reset_ports : [
  {"name" : "reset_io", "active_state" : "1"},
  {"name" : "jtag_reset", "active_state" : "1"}
]

# BSD and JTAG
synthesis.insert_dft.bsd : false 
synthesis.dft.jtag.clk : "jtag_TCK"
synthesis.dft.jtag.tdi : "jtag_TDI"
synthesis.dft.jtag.tdo : "jtag_TDO"
synthesis.dft.jtag.tms : "jtag_TMS"
synthesis.dft.jtag.reset : "jtag_reset"

# Memory Collar configuration 
synthesis.insert_dft.memory_wrapper : true 

    \end{lstlisting}
    \end{minipage}
    \caption{High-Level Hammer YAML configuration for automated DfT insertion.}
    \label{fig:hammer_dft_config}
\end{figure}

Figure~\ref{fig:hammer_dft_config} shows a representative YAML
configuration for the DfT insertion step during the synthesis action, illustrating how scan
chain count, flip-flop style, port mapping, and memory collar can all be controlled from a single high-level
configuration file without interacting directly with the EDA tool. It is divided into four main functional blocks:

\begin{itemize}
    \item \textbf{Global DfT Master Switch (line 2):}
    Acts as the top-level enable flag for the DfT insertion engine in the synthesis. When enabled, the tool executes DfT insertion based on the other parameters.

    \item \textbf{Internal Scan Chain Insertion (line 5-25):}
    Configures structural scan path insertion to replace standard sequential elements with scan equivalents and stitch them together:

    \begin{itemize}
        \item \texttt{style}: Dictates the cell architecture used to make sequential elements scannable (e.g., multiplexed flip-flops vs. clocked scan).
        \item \texttt{chain\_count}: Controls parallelism in the scan path by specifying how many distinct scan chains to construct across the flip-flops.
        \item \texttt{ports}: Directs the tool to route the main scan input and output lines to explicit top-level primary IO pins.
        \item \texttt{clock\_ports}: Identifies which clocks must remain driven and controllable during shift and capture modes, alongside their active states and waveform timing windows for test pattern generation.
        \item \texttt{reset\_ports}: Informs the synthesis tool of asynchronous or synchronous reset signals that need to be driven into known, inactive, or controllable states during test pattern shifting to avoid unintended resets.
    \end{itemize}

    \item \textbf{Boundary Scan and TAP Controller Integration (line 28-33):}
    Handles IEEE 1149.1 Standard boundary scan and JTAG Test Access Port (TAP) setup:
    \begin{itemize}
        \item \texttt{insert\_dft.bsd}: Toggles whether the synthesis tool automatically generates boundary scan cells around primary I/O pads for board-level interconnect testing.
        \item \texttt{jtag.* mappings}: Binds the TAP state machine’s standard interface signals (Test Clock, Test Data In/Out, Test Mode Select, and Test Reset) to the designated top-level signals.
    \end{itemize}

    \item \textbf{Memory Test \& Wrapper Configuration (line 36):}
    Toggles automatic memory collar insertion. When active, it wraps memory instances with test wrappers (or Memory BIST interfaces) to allow memory arrays to be accessed, controlled, bypassed, and tested through the scan network.
\end{itemize}

\subsubsection{Automatic Test Pattern Generation}
The automatic test pattern generation action is in charge of generating the test patterns applied by the ATE used during the manufacturing test flow to test the manufactured silicon devices. This action integrates the ATPG capabilities into the Hammer environment, providing a standardized interface to ATPG tools.

\begin{figure}[!htbp]
    \centering
    \includegraphics[width=1\columnwidth]{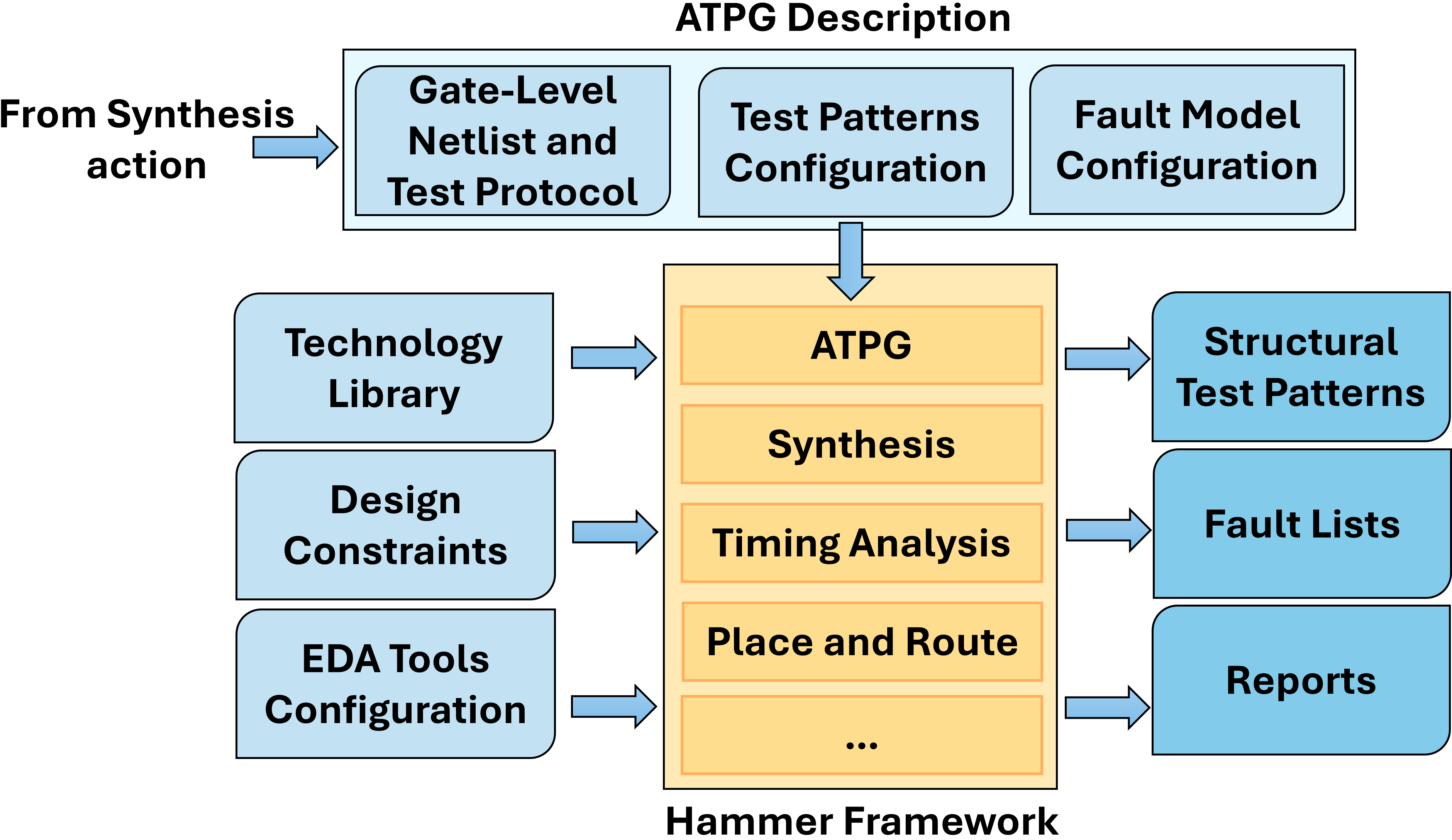}
    \caption{ATPG action in Hammer framework.}
    \label{fig:atpg}
\end{figure}

Figure~\ref{fig:atpg} shows the data flow for the added ATPG action.
In order to effectively use the ATPG action, three input data categories must be provided:
\begin{itemize}
    \item \textbf{Design data:} The gate-level netlist (or submodules) with the DfT architecture and test protocol.
    \item \textbf{Fault modeling:} The fault model description, i.e., logical models for physical defects, such as stuck-at faults, transition delay faults, small delay faults, bridge faults.
    \item \textbf{Tool configurations:} Users can specify custom arguments for the ATPG tool (they may be tool-dependent), such as test protocol consistency check, elaboration of synthesized design, the ATPG effort limit (i.e., the time and/or algorithm-dependent metrics the ATPG engine invests in generating a test pattern), maximum test patterns generated and test coverage goal, directly through the configuration files. This enables a fine-grained control and flexibility over the pattern generation process.

\end{itemize}

\begin{figure}[!htbp]
    \centering
    \begin{minipage}{0.9\columnwidth}
    \begin{lstlisting}[style=hammer_yaml]
atpg.target_coverage : 98.5
atpg.max_test_patterns : 0 .
# Args for the Design rule check (DRC)
atpg.<tool>.set_drc_args: [""]
# Args for the DRC execution
# (spf file is automatically taken after synthesis)
atpg.<tool>.run_drc_args: [""]
# Args for the ATPG configuration
atpg.<tool>.set_atpg_args: [""]
# Args for the ATPG execution
atpg.<tool>.run_atpg_args: ["-auto_compression"]

# Args for the Fault generation
atpg.<tool>.add_faults_args: ["-all"]
    \end{lstlisting}
    \end{minipage}
    \caption{Hammer YAML configuration for the ATPG action,
    exposing per-command arguments for DRC, pattern generation,
    fault simulation, and fault model setup. \texttt{<tool>} can be
    replaced by the actual tool name at configuration time.}
    \label{fig:hammer_atpg_config}
\end{figure}

Figure~\ref{fig:hammer_atpg_config} shows a representative YAML
configuration for the ATPG action, illustrating how per-command
arguments for design rule checking, pattern generation, fault
simulation, and fault model setup can be specified directly in
the high-level configuration file and forwarded to the underlying
ATPG tool without manual script adaptation.

The ATPG action outputs include structural test patterns (in stil format~\cite{IEEE_STIL} for being used by the ATEs, there is also the possibility to specify another format if necessary), the fault model-dependent fault list with each fault marked as detected or not by the test patterns, coverage reports and additional reports such as the analysis of ATPG untestable faults. Notably, the implementation supports the \textit{re-run} capability, allowing users to tune ATPG-specific parameters or perform iterative generation processes.

\subsubsection{Functional Fault Simulation}
The functional fault simulation action is in charge of evaluating the fault detection capabilities of functional test programs—such as Software-Based Self-Test (SBST) or System-Level Test (SLT) on the gate-level netlist of the synthesized circuit. This evaluation determines whether functional programs can capture additional faults compared to structural tests (for SLT during manufacturing) or evaluate coverage against a customized fault list (for online testing).

\begin{figure}[!htbp]
\centering
\includegraphics[width=1\columnwidth]{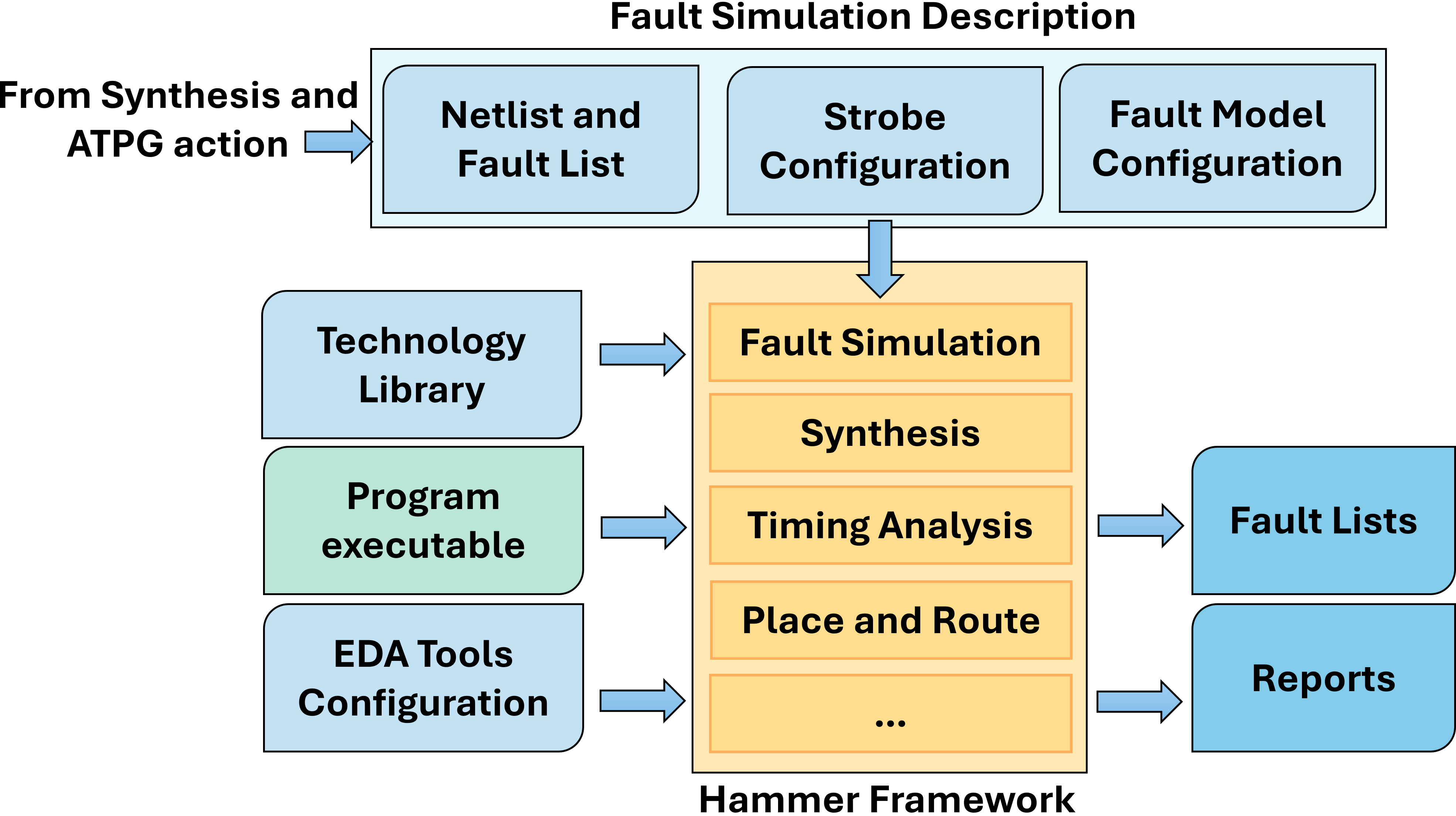}
\caption{Fault Simulation action in Hammer framework.}
\label{fig:fsim}
\end{figure}

Figure~\ref{fig:fsim} shows the data flow for the added functional fault simulation action, which fault simulates a target program executable and assesses its fault detection capabilities.

In order to effectively use the functional fault simulation action, four input data categories must be provided:
\begin{itemize}
    \item \textbf{Design \& Test Executable data:} The gate-level netlist and the target functional test program executable.
    \item \textbf{Fault modeling:} The fault model description, i.e., logical models for physical defects, such as stuck-at faults, transition delay faults, small delay faults, and bridge faults.
    \item \textbf{Fault scope \& target lists:} An optional manufacturing fault list (e.g., imported from prior ATPG runs) to perform incremental fault simulation, or target fault locations from which new faults are generated.
    \item \textbf{Strobe \& observation configuration:} The specification of when and where the behaviors of the good machine and the faulty machine are observed and compared to promote a fault to a detected status.
\end{itemize}

Figure~\ref{fig:hammer_fsim_config} shows a representative YAML configuration for the functional fault simulation action, illustrating how simulation parameters, coverage metrics, and fault injection scoping can be specified directly in the high-level configuration file:
\begin{itemize}
    \item \textbf{Elaboration options:} The \textit{optional\_elaboration\_flags} parameter allows users to pass custom command-line arguments to the underlying simulator during design hierarchy elaboration.
    \item \textbf{Custom coverage metrics:} The \textit{weights}, \textit{test\_coverage}, and \textit{fault\_coverage} entries enable user-defined formulas and status multipliers (e.g., assigning a standard 1.0 weight to potentially detected faults, \textit{PT}) for mathematical scoring.
    \item \textbf{Fault location scoping:} The \textit{fault\_locations} block defines injection regions using regular expressions (e.g., \textit{'**'} to target a module alongside all internal sub-modules), specifying fault types, location primitive types (e.g., \textit{PORT}), and exclusion flags.
    \item \textbf{Observation points:} The \textit{strobe\_modules} section defines target modules or signals where good and faulty machine behaviors are strobed and compared.
\end{itemize}

\begin{figure}[!htbp]
    \centering
    \begin{minipage}{0.9\columnwidth}
    \begin{lstlisting}[style=hammer_yaml]
fsim.inputs:

  # Optional elaboration flags for the fault simulator
  optional_elaboration_flags: [""]

  # Optional execution flags for the fault simulator
  optional_runtime_flags: [""]

  # Fault locations and types to inject
  fault_locations: [
    {location: "ChipTop.**",
     fault_type: "R,F",
     type_of_fault_location: "PORT",
     exclude: False},
  ]

  # Strobe modules/signals for good/faulty machine comparison
  strobe_modules: [
    "`TOPLEVEL.core",
  ]
    \end{lstlisting}
    \end{minipage}
    \caption{Hammer YAML configuration for the functional fault
    simulation action, covering elaboration flags, coverage formula
    weights, fault location scoping, and strobe module specification.}
    \label{fig:hammer_fsim_config}
\end{figure}

The functional fault simulation action outputs include updated fault lists with detection statuses, hierarchical coverage reports, and detailed reports on undetected faults. Notably, the implementation supports incremental fault simulation via user-defined fault lists (e.g., originating from the ATPG action) and accepts custom user-defined strobe files, providing fine-grained control over the observation environment and fault detection logic.

\section{Experimental results}
\label{sec:experimental_results}

This section presents the first comprehensive structural and functional
test results obtained by using the \textit{Testyard} framework.
The experimental campaign is organized to validate the three newly
introduced actions (DfT insertion, ATPG, and functional fault
simulation) across multiple SoC configurations (RISC-V based) and technology nodes.

\subsection{Experimental setup}
The Register-Transfer Level (RTL) designs of the case studies, generated by the Chipyard framework, are synthesized using different open-source cell libraries, and the scan chain is inserted using a commercial tool.

The structural tests are generated and fault simulated using two different commercial tools from two different vendors, for fault models such as stuck-at, transition delay, and small delay faults. For small delay fault models, a commercial timing analysis tool is used to perform static timing analysis for correctly generating the input timing data required by the commercial ATPG tools.

Functional tests are first assessed using the \textit{Spike} ISA simulator and then logically simulated using commercial and open-source simulators. The functional fault simulations are executed using a commercial fault simulator tool for various fault models, including stuck-at, transition delay, and small delay faults.

All experiments and computations have been executed on a high-performance multi-processor server equipped with a 64-bit 112-core \textit{Intel Xeon Gold 6238R} CPU at 2.20~GHz, 256~GB of RAM, a storage system of 10~TBytes, and a \textit{Rocky Linux 8} operating system.

Table~\ref{tab:tech_libraries} summarizes the technology libraries employed for the synthesis of the case studies. The technology libraries span three nodes: the
 45$nm$ Planar node, the 15$nm$ FinFET node, and the advanced 7$nm$ FinFET node. They include industry-standard academic libraries like ASAP7 (7nm FinFET) and Nangate (15$nm$ and 45$nm$).

\begin{table}[!htbp]
    \centering
    \caption{Overview of evaluated technology libraries.}
    \label{tab:tech_libraries}
    \begin{tabular}{|c|c|c|c|}
        \hline
        \textbf{Name} & \textbf{Node} & \textbf{Device Type} & \textbf{Sourcing Entity} \\
        \hline
        \hline
        ASAP7 & 7$nm$ & FinFET & ASU/Arm\cite{ASAP7} \\
        NANGATE & 15$nm$ & FinFET & Silvaco\cite{SILVACO_OPENPDK} \\
        NANGATE & 45$nm$ & Planar MOSFETs & Silvaco \cite{SILVACO_OPENPDK} \\
        \hline
    \end{tabular}
\end{table}
This variety allows for a comprehensive analysis across different transistor technologies and scaling generations.

\subsection{Design-for-Testability Description}

\begin{table*}[!t]
    \centering
    \caption{Post-DfT cell count, flip-flop (scan cell and non) count, and area overhead introduced by DfT insertion, per case study and technology library.}
    \label{tab:dft_area_overhead}
    \setlength{\tabcolsep}{4pt}
    \renewcommand{\arraystretch}{1.0}
    \resizebox{\textwidth}{!}{%
    \begin{tabular}{|l|r|r|r|r|r|r|r|r|r|}
        \hline
        \multirow{2}{*}{\textbf{Config Name}} &
          \multicolumn{3}{c|}{\textbf{Nangate45}} &
          \multicolumn{3}{c|}{\textbf{Nangate15}} &
          \multicolumn{3}{c|}{\textbf{ASAP7}} \\ \cline{2-10}
        & \textbf{\#cells} & \textbf{\#FFs} & \textbf{Overhead [\%]} & \textbf{\#cells} & \textbf{\#FFs} & \textbf{Overhead [\%]} & \textbf{\#cells} & \textbf{\#FFs} & \textbf{Overhead [\%]} \\
        \hline
        \hline
         AES256ECBRocketConfig           & 912,450   & 157,873   & 6.3  & 1,001,539 & 157,873   & 6.3  & 744,374   & 160,798   & 19.1  \\ \hline
        CVA6Config                      & 348,366   & 56,396    & 13.4 & 408,647   & 56,461    & 13.3 & 297,986   & 56,073    & 12.5 \\ \hline
        MediumBoomV3Config              & 771,252   & 142,149   & 16.4 & 1,033,139 & 142,148   & 16.2 & 730,766   & 144,472   & 32.6 \\ \hline
        MediumBoomV4Config              & 781,043   & 157,261   & 13.3 & 1,094,566 & 157,260   & 14.3 & 758,862   & 159,670   & 33.3 \\ \hline
        GemminiRocketConfig             & 1,234,336 & 158,965   & 11.6 & 1,576,885 & 158,965   & 13.3 & 1,243,545 & 161,865   & 8.8  \\ \hline
        IbexConfig                      & 127,354   & 26,270    & 18.0 & 175,796   & 26,614    & 25.6 & 104,655   & 26,576    & 12.8 \\ \hline
        FFTRocketConfig                 & 264,152   & 44,820    & 8.3  & 343,162   & 44,820    & 8.2  & 249,065   & 45,863    & 23.0 \\ \hline
        ShuttleConfig                   & 350,622   & 62,761    & 9.3  & 456,702   & 62,760    & 8.4  & 315,700   & 63,973    & 19.8 \\ \hline
        MultiNoCConfig                  & 4,274,657 & 595,656   & 8.6  & 5,291,361 & 596,048   & 9.0  & 3,750,555 & 607,706   & 17.7 \\ \hline
        RocketConfig                    & 257,407   & 44,051    & 8.2  & 335,153   & 44,051    & 8.1  & 236,032   & 45,078    & 20.2 \\ \hline
        ManyPeripheralsRocketConfig           & 334,095   & 60,107    & 8.6  & 419,290   & 58,543    & 6.0  & 346,085   & 58,201    &  22.9 \\ \hline
        REFV256D128M64RocketConfig      & 694,492   & 77,528    & 5.5  & 838,195   & 77,528    & 3.3  & 664,069   & 79,102    & 12.9 \\ \hline
        DualRocketConfig                & 398,701   & 62,144    & 6.7  & 511,514   & 62,145    & 5.9  & 368,197   & 63,395    & 17.1 \\ \hline
    \end{tabular}%
    }
\end{table*}

The Design-for-Testability (DfT) insertion action described in Section~\ref{sec:hammer} was applied uniformly across all synthesized benchmarks listed in Table~\ref{tab:case_studies_architectural_description}, adopting the same scan architecture. This choice keeps the structural test infrastructure orthogonal to the specific SoC configuration, so that the testing effort reported in Section~\ref{sec:experimental_results} can be attributed to the design's own complexity rather than to differences in the scan implementation across benchmarks.

\subsubsection{Scan configuration}

All benchmarks adopt a \textit{multiplexed flip-flop} (mux-scan) style, with the associated scan-enable and test-mode control ports (\texttt{test\_se}, \texttt{test\_mode}), in which the data input of every functional flip-flop is augmented with a two-input multiplexer selecting between the functional data path and the scan input during the shift phase, consistent with the configuration excerpt already introduced in Figure~\ref{fig:hammer_dft_config}. The scan chains are attached to the JTAG Test Access Port already present in every case study, reusing the \texttt{jtag\_TDI}/\texttt{jtag\_TDO} ports as chain input/output and \texttt{jtag\_TCK} as the shift clock, while the remaining functional clocks (e.g., \texttt{clock\_uncore}, \texttt{serial\_tl\_0\_clock\_in}) are treated as independent capture domains during ATPG. The scan-enable, scan-clock, and chain-serial networks are fully routed at the synthesis stage under a high DfT mapping effort, and the result is verified with a dedicated DfT design-rule check (DRC) pass before and after insertion.

In addition to scan-chain insertion, memory collar test-point insertion is enabled by default: every SRAM macro identified through the technology plugin's macro library is instrumented with observe points at its data-input pins and control points at its data-output pins through a shadow-wrapper test-point analysis pass.

Table~\ref{tab:dft_config} summarizes the scan configuration parameters shared by all case studies.

\begin{table}[!htbp]
    \centering
    \caption{Scan configuration parameters shared by all synthesized benchmarks.}
    \label{tab:dft_config}
    \begin{tabular}{|l|c|}
        \hline
        \textbf{Parameter} & \textbf{Value} \\
        \hline
        \hline
        Scan cell style & Multiplexed flip-flop \\
        \hline
        Scan enable / test-mode ports & \texttt{test\_se}, \texttt{test\_mode} (active-high) \\
        \hline
        Number of scan chains & $1$ \\
        \hline
        Clock mixing across chains & Enabled \\
        \hline
        Scan shift clock & \texttt{jtag\_TCK} \\
        \hline
        Scan data I/O & \texttt{jtag\_TDI} / \texttt{jtag\_TDO} \\
        \hline
        Memory collar insertion & Enabled \\
        \hline
        Scan compression & Disabled \\
        \hline
    \end{tabular}
\end{table}
Although single scan chains are rarely used in complex designs or for testing delay faults, we adopted a single-chain setup as a baseline DfT architecture to facilitate future enhancements.

\subsubsection{Area overhead}

Inserting the scan architecture described in the previous subsection is not area-neutral: converting a functional flip-flop into a scan cell requires an additional two-input multiplexer on its data input, and routing the resulting scan chains introduces incremental buffering to meet the balancing and timing constraints of the synthesis step. Beyond the cell-level contribution, the DfT insertion action also instantiates auxiliary test infrastructure, such as the TAP controller logic driving the JTAG-based scan interface and, where enabled, the memory collars wrapping SRAM macros, all of which add to the post-synthesis cell area relative to a functionally equivalent, non-scan design.

Furthermore, the relative area cost of a scan multiplexer with respect to a plain flip-flop is technology-dependent: the same scan configuration can therefore result in a different percentage overhead when synthesized on Nangate45, Nangate15 and ASAP7 libraries listed in Table~\ref{tab:tech_libraries}, reflecting differences in standard-cell characterization rather than in the DfT methodology itself.

Table~\ref{tab:dft_area_overhead} reports the area overhead introduced by DfT insertion for each case study across the three evaluated technology libraries, following the same benchmark ordering used in Table~\ref{tab:case_studies_architectural_description} to ease cross-reference.
\begin{table*}[p]
\centering
\caption{ATPG experimental results for different fault models across the three evaluated technology libraries. `\#Faults' reports the total number of targeted faults, shared across the three fault models; `UT' is the percentage of ATPG untestable faults over the total fault count, `TC' denotes the achieved test generation coverage, and `\#Patterns' reports the number of generated test patterns.}
\label{tab:atpg_all}
\setlength{\tabcolsep}{2.2pt}
\renewcommand{\arraystretch}{1.05}

\subfloat[Nangate45]{%
\resizebox{\textwidth}{!}{%
\begin{tabular}{|l|r|r|r|r|r|r|r|r|r|r|}
\hline

\multicolumn{1}{|c|}{\multirow{3}{*}{\textbf{Config Name}}} &
\multicolumn{10}{c|}{\textbf{ATPG results - Nangate45}} \\ \cline{2-11}

\multicolumn{1}{|c|}{} &
\multicolumn{1}{c|}{\multirow{2}{*}{\textbf{\#Faults}}} &
\multicolumn{3}{c|}{\textbf{SAF}} &
\multicolumn{3}{c|}{\textbf{TDF}} &
\multicolumn{3}{c|}{\textbf{SDF}} \\ \cline{3-11}

\multicolumn{1}{|c|}{} &
\multicolumn{1}{c|}{} &
\multicolumn{1}{c|}{\textbf{UT [\%]}} &
\multicolumn{1}{c|}{\textbf{TC [\%]}} &
\multicolumn{1}{c|}{\textbf{\#Patterns}} &
\multicolumn{1}{c|}{\textbf{UT [\%]}} &
\multicolumn{1}{c|}{\textbf{TC [\%]}} &
\multicolumn{1}{c|}{\textbf{\#Patterns}} &
\multicolumn{1}{c|}{\textbf{UT [\%]}} &
\multicolumn{1}{c|}{\textbf{TC [\%]}} &
\multicolumn{1}{c|}{\textbf{\#Patterns}} \\ \hline \hline

AES256ECBRocketConfig & 7,354,394 & 4.08 & 95.60 & 2,289 & 4.44 & 95.22 & 5,436 & 4.58 & 75.26 & 9,330 \\
\hline
CVA6Config & 2,650,648 & 3.16 & 96.63 & 4,420 & 4.27 & 95.08 & 9,613 & 4.38 & 72.09 & 13,667 \\
\hline
MediumBoomV3Config & 6,558,356 & 3.19 & 96.63 & 4,796 & 4.94 & 94.75 & 4,511 & 5.00 & 76.39 & 10,310 \\
\hline
MediumBoomV4Config & 6,957,208 & 2.52 & 97.30 & 3,745 & 6.11 & 93.62 & 3,663 & 6.08 & 68.74 & 8,349 \\
\hline
GemminiRocketConfig & 9,955,836 & 3.10 & 93.60 & 3,148 & 3.48 & 92.35 & 5,959 & 3.63 & 29.19 & 8,095 \\
\hline
IbexConfig & 1,049,732 & 12.69 & 86.82 & 796 & 15.18 & 84.23 & 1,222 & 15.10 & 55.19 & 1,942 \\
\hline
FFTRocketConfig & 2,158,410 & 5.99 & 93.39 & 2,315 & 7.47 & 91.94 & 3,855 & 8.06 & 72.06 & 6,777 \\
\hline
ShuttleConfig & 2,948,170 & 5.67 & 93.87 & 2,738 & 7.19 & 92.37 & 3,430 & 7.23 & 69.10 & 6,854 \\
\hline
MultiNoCConfig & 36,822,786 & 1.03 & 98.54 & 840 & 1.02 & 98.50 & 1,979 & 1.21 & 75.42 & 5,257 \\
\hline
RocketConfig & 2,108,074 & 6.25 & 93.17 & 2,273 & 7.65 & 91.75 & 3,918 & 8.24 & 72.43 & 6,336 \\
\hline
ManyPeripheralsRocketConfig & 2,653,752 & 4.93 & 94.60 & 2,141 & 6.30 & 93.23 & 4,065 & 6.70 & 71.93 & 6,499 \\
\hline
REFV256D128M64RocketConfig & 5,391,968 & 2.52 & 97.08 & 5,707 & 3.13 & 96.29 & 9,011 & 3.42 & 62.68 & 15,084 \\
\hline
DualRocketConfig & 3,225,532 & 4.51 & 94.95 & 2,502 & 5.60 & 93.83 & 4,100 & 6.42 & 71.45 & 6,397 \\
\hline
\end{tabular}%
}
\label{tab:atpg_nangate45}
}

\vspace{1em}

\subfloat[Nangate15]{%
\resizebox{\textwidth}{!}{%
\begin{tabular}{|l|r|r|r|r|r|r|r|r|r|r|}
\hline

\multicolumn{1}{|c|}{\multirow{3}{*}{\textbf{Config Name}}} &
\multicolumn{10}{c|}{\textbf{ATPG results - Nangate15}} \\ \cline{2-11}

\multicolumn{1}{|c|}{} &
\multicolumn{1}{c|}{\multirow{2}{*}{\textbf{\#Faults}}} &
\multicolumn{3}{c|}{\textbf{SAF}} &
\multicolumn{3}{c|}{\textbf{TDF}} &
\multicolumn{3}{c|}{\textbf{SDF}} \\ \cline{3-11}

\multicolumn{1}{|c|}{} &
\multicolumn{1}{c|}{} &
\multicolumn{1}{c|}{\textbf{UT [\%]}} &
\multicolumn{1}{c|}{\textbf{TC [\%]}} &
\multicolumn{1}{c|}{\textbf{\#Patterns}} &
\multicolumn{1}{c|}{\textbf{UT [\%]}} &
\multicolumn{1}{c|}{\textbf{TC [\%]}} &
\multicolumn{1}{c|}{\textbf{\#Patterns}} &
\multicolumn{1}{c|}{\textbf{UT [\%]}} &
\multicolumn{1}{c|}{\textbf{TC [\%]}} &
\multicolumn{1}{c|}{\textbf{\#Patterns}} \\ \hline \hline

AES256ECBRocketConfig & 7,702,008 & 4.29 & 95.37 & 2,532 & 4.67 & 94.79 & 5,291 & 4.77 & 73.98 & 11,220 \\
\hline
CVA6Config & 2,702,382 & 4.12 & 95.51 & 4,542 & 5.40 & 93.94 & 9,397 & 5.44 & 73.79 & 12,802 \\
\hline
MediumBoomV3Config & 7,022,048 & 3.38 & 96.40 & 3,983 & 4.74 & 94.74 & 4,208 & 4.76 & 78.68 & 7,391 \\
\hline
MediumBoomV4Config & 7,566,686 & 4.02 & 95.63 & 3,636 & 6.20 & 93.06 & 4,195 & 6.21 & 68.24 & 7,570 \\
\hline
GemminiRocketConfig & 10,681,356 & 3.18 & 93.56 & 2,966 & 3.40 & 92.78 & 4,838 & 3.69 & 30.22 & 8,742 \\
\hline
IbexConfig & 1,149,276 & 16.31 & 82.85 & 649 & 17.76 & 80.88 & 1,154 & 17.79 & 60.46 & 1,795 \\
\hline
FFTRocketConfig & 2,294,346 & 7.11 & 92.29 & 2,175 & 8.06 & 91.05 & 3,718 & 8.35 & 70.47 & 6,127 \\
\hline
ShuttleConfig & 3,105,074 & 6.41 & 93.12 & 3,084 & 7.45 & 91.82 & 3,264 & 7.48 & 67.96 & 6,313 \\
\hline
MultiNoCConfig & 38,532,222 & 1.04 & 98.53 & 840 & 1.07 & 98.50 & 2,047 & 1.21 & 66.59 & 5,001 \\
\hline
RocketConfig & 2,242,062 & 7.28 & 92.12 & 2,124 & 8.22 & 90.89 & 3,721 & 8.53 & 72.29 & 6,489 \\
\hline
ManyPeripheralsRocketConfig & 2,792,874 & 5.80 & 93.70 & 2,113 & 6.83 & 92.43 & 3,824 & 7.06 & 71.60 & 6,421 \\
\hline
REFV256D128M64RocketConfig & 5,617,026 & 3.09 & 96.55 & 5,435 & 3.51 & 95.85 & 9,019 & 3.69 & 59.34 & 14,887 \\
\hline
DualRocketConfig & 3,429,820 & 5.23 & 94.26 & 2,490 & 6.03 & 93.22 & 3,763 & 6.41 & 68.82 & 6,580 \\
\hline
\end{tabular}%
}
\label{tab:atpg_nangate15}
}

\vspace{1em}

\subfloat[ASAP7]{%
\resizebox{\textwidth}{!}{%
\begin{tabular}{|l|r|r|r|r|r|r|r|r|r|r|}
\hline

\multicolumn{1}{|c|}{\multirow{3}{*}{\textbf{Config Name}}} &
\multicolumn{10}{c|}{\textbf{ATPG results - ASAP7}} \\ \cline{2-11}

\multicolumn{1}{|c|}{} &
\multicolumn{1}{c|}{\multirow{2}{*}{\textbf{\#Faults}}} &
\multicolumn{3}{c|}{\textbf{SAF}} &
\multicolumn{3}{c|}{\textbf{TDF}} &
\multicolumn{3}{c|}{\textbf{SDF}} \\ \cline{3-11}

\multicolumn{1}{|c|}{} &
\multicolumn{1}{c|}{} &
\multicolumn{1}{c|}{\textbf{UT [\%]}} &
\multicolumn{1}{c|}{\textbf{TC [\%]}} &
\multicolumn{1}{c|}{\textbf{\#Patterns}} &
\multicolumn{1}{c|}{\textbf{UT [\%]}} &
\multicolumn{1}{c|}{\textbf{TC [\%]}} &
\multicolumn{1}{c|}{\textbf{\#Patterns}} &
\multicolumn{1}{c|}{\textbf{UT [\%]}} &
\multicolumn{1}{c|}{\textbf{TC [\%]}} &
\multicolumn{1}{c|}{\textbf{\#Patterns}} \\ \hline \hline

AES256ECBRocketConfig & 5,467,150 & 5.75 & 92.98 & 2,334 & 7.59 & 90.59 & 3,998 & 7.84 & 66.29 & 5,613 \\
\hline
CVA6Config & 2,120,004 & 58.56 & 31.97 & 789 & 58.87 & 30.47 & 1,137 & 58.93 & 12.36 & 1,722 \\
\hline
MediumBoomV3Config & 5,482,378 & 7.12 & 90.88 & 3,736 & 11.19 & 85.74 & 3,524 & 11.28 & 64.81 & 5,768 \\
\hline
MediumBoomV4Config & 5,851,724 & 5.30 & 93.28 & 2,601 & 9.78 & 88.13 & 3,330 & 9.80 & 69.24 & 8,021 \\
\hline
GemminiRocketConfig & 9,047,954 & 3.02 & 90.39 & 3,455 & 4.23 & 88.49 & 7,510 & 4.91 & 29.54 & 10,401 \\
\hline
IbexConfig & 759,366 & 25.77 & 70.43 & 822 & 29.49 & 67.04 & 1,194 & 29.42 & 46.29 & 1,608 \\
\hline
FFTRocketConfig & 1,800,798 & 9.56 & 86.59 & 1,805 & 12.75 & 82.78 & 3,050 & 13.18 & 58.65 & 5,097 \\
\hline
ShuttleConfig & 2,360,872 & 8.31 & 90.01 & 1,748 & 11.37 & 86.42 & 3,351 & 11.49 & 66.27 & 5,844 \\
\hline
MultiNoCConfig & 30,425,280 & 2.73 & 96.46 & 2,210 & 3.96 & 94.80 & 4,339 & 4.23 & 72.50 & 8,135 \\
\hline
RocketConfig & 1,720,516 & 9.78 & 88.09 & 1,716 & 12.87 & 84.28 & 2,942 & 13.46 & 63.38 & 4,356 \\
\hline
ManyPeripheralsRocketConfig & 2,623,452 & 8.18 & 90.57 & 1,822 & 10.77 & 87.49 & 3303 & 11.27 & 60.80 & 8483 \\
\hline
REFV256D128M64RocketConfig & 4,771,272 & 3.91 & 95.14 & 4,869 & 5.53 & 91.70 & 7,315 & 6.07 & 41.14 & 8,686 \\
\hline
DualRocketConfig & 2,675,988 & 7.88 & 90.37 & 1,833 & 10.81 & 86.46 & 3,190 & 11.37 & 59.50 & 6,809 \\
\hline
\end{tabular}%
}
\label{tab:atpg_asap7}
}

\end{table*}

\begin{table*}[t]
\centering
\caption{Scan patterns fault simulation experimental results for different fault models on the Nangate15 technology library, obtained by re-simulating the generated test patterns with a second commercial ATPG tool from a different vendor. `\#Faults' reports the total number of targeted faults, shared across the three fault models; `UT' is the percentage of untestable faults over the total fault count, `TC' denotes the achieved test coverage after fault simulation, and `\#Patterns' reports the number of simulated test patterns.}
\label{tab:atpg_fault_simulation_15}
\setlength{\tabcolsep}{2.2pt}
\renewcommand{\arraystretch}{1.05}

\resizebox{\textwidth}{!}{%
\begin{tabular}{|l|r|r|r|r|r|r|r|r|r|r|}
\hline

\multicolumn{1}{|c|}{\multirow{3}{*}{\textbf{Config Name}}} &
\multicolumn{10}{c|}{\textbf{Scan patterns fault simulation results - Nangate15}} \\ \cline{2-11}

\multicolumn{1}{|c|}{} &
\multicolumn{1}{c|}{\multirow{2}{*}{\textbf{\#Faults}}} &
\multicolumn{3}{c|}{\textbf{SAF}} &
\multicolumn{3}{c|}{\textbf{TDF}} &
\multicolumn{3}{c|}{\textbf{SDF}} \\ \cline{3-11}

\multicolumn{1}{|c|}{} &
\multicolumn{1}{c|}{} &
\multicolumn{1}{c|}{\textbf{UT [\%]}} &
\multicolumn{1}{c|}{\textbf{TC [\%]}} &
\multicolumn{1}{c|}{\textbf{\#Patterns}} &
\multicolumn{1}{c|}{\textbf{UT [\%]}} &
\multicolumn{1}{c|}{\textbf{TC [\%]}} &
\multicolumn{1}{c|}{\textbf{\#Patterns}} &
\multicolumn{1}{c|}{\textbf{UT [\%]}} &
\multicolumn{1}{c|}{\textbf{TC [\%]}} &
\multicolumn{1}{c|}{\textbf{\#Patterns}} \\ \hline \hline

AES256ECBRocketConfig & 7,701,926 & 3.24 & 94.81 & 2,531 & 3.81 & 94.29 & 5,290 & 4.26 & 93.61 & 11,219 \\
\hline
CVA6Config & 2,837,186 & 3.44 & 92.91 & 4,541 & 4.34 & 91.11 & 9,396 & 4.34 & 79.87 & 12,801 \\
\hline
MediumBoomV3Config & 7,022,048 & 2.80 & 96.15 & 3,982 & 3.28 & 95.00 & 4,207 & 4.14 & 93.92 & 7,390 \\
\hline
MediumBoomV4Config & 7,566,686 & 2.44 & 95.31 & 3,635 & 2.83 & 93.09 & 4,194 & 4.67 & 92.16 & 7,569 \\
\hline
GemminiRocketConfig & 10,681,356 & 2.72 & 92.58 & 2,965 & 3.38 & 91.62 & 4,837 & 3.38 & 89.96 & 8,741 \\
\hline
IbexConfig & 1,149,276 & 13.50 & 82.35 & 648 & 14.74 & 80.99 & 1,153 & 14.74 & 76.65 & 1,794 \\
\hline
FFTRocketConfig & 2,294,346 & 6.93 & 91.64 & 2,174 & 7.86 & 90.61 & 3,717 & 8.36 & 88.99 & 6,126 \\
\hline
ShuttleConfig & 3,105,074 & 5.89 & 92.62 & 3,083 & 6.94 & 91.59 & 3,263 & 7.40 & 90.52 & 6,312 \\
\hline
MultiNoCConfig & 38,532,222 & 0.90 & 98.22 & 839 & 1.02 & 98.23 & 2,046 & 1.18 & 84.84 & 5,000 \\
\hline
RocketConfig & 2,242,062 & 7.10 & 91.46 & 2,123 & 8.09 & 90.40 & 3,720 & 8.56 & 88.90 & 6,488 \\
\hline
ManyPeripheralsRocketConfig & 2,792,874 & 5.68 & 93.13 & 2,112 & 6.43 & 92.19 & 3,823 & 6.86 & 88.98 & 6,420 \\
\hline
REFV256D128M64RocketConfig & 5,617,026 & 2.88 & 95.14 & 5,434 & 3.27 & 94.49 & 9,018 & 3.48 & 93.79 & 14,886 \\
\hline
DualRocketConfig & 3,429,820 & 5.27 & 93.58 & 2,489 & 5.91 & 92.71 & 3,762 & 6.53 & 91.45 & 6,579 \\
\hline
\end{tabular}%
}

\end{table*}

\subsection{Structural test evaluation}

This subsection reports the structural test results obtained within the \textit{Testyard} framework. Section~\ref{sec:atpg_methodology} describes the fault-model configuration shared across all experiments, while Section~\ref{sec:atpg_results} presents the achieved fault coverage and pattern counts for the case studies of Table~\ref{tab:case_studies_architectural_description}, across the three technology libraries of Table~\ref{tab:tech_libraries} and the three fault models (SAF, TDF and SDF) introduced in Section~\ref{sec:hammer}.

\subsubsection{ATPG Methodology}
\label{sec:atpg_methodology}

Pattern generation runs with automatic pattern compaction enabled for every fault model, and the complete fault list is targeted in each single run. The number of generated test patterns is unconstrained, while the test coverage target is set to 99.00\% with medium ATPG effort.

The transition delay and small delay faults are both generated under the same underlying transition fault model: small delay ATPG additionally loads the static timing slack data and credits a fault only when it is detected on its own minimum-slack path, whereas transition delay ATPG applies no such constraint. Since both share the same underlying fault list, transition delay and small delay results in Tables~\ref{tab:atpg_nangate45}--\ref{tab:atpg_asap7} report the same fault count for a given benchmark and technology library, differing only in achieved coverage and pattern count as a consequence of the additional timing information.

Every fault targeted by ATPG is ultimately classified as detected, undetectable (provably impossible to activate or observe through the available scan access), or ATPG untestable (no detecting pattern could be found by the tool within the allotted effort, without a proof of redundancy). The `UT' column reported in Tables~\ref{tab:atpg_nangate45}--\ref{tab:atpg_asap7} refers specifically to this latter category, expressed as a percentage of the total fault count for the corresponding fault model, and does not include the separate population of provably undetectable faults.

The ATPG campaign presented in this work targets a baseline, single-pass structural test configuration and does not explore several advanced capabilities supported by commercial ATPG tools. Test data compression is disabled for every benchmark (Table~\ref{tab:dft_config}), consistent with the single, uncompressed scan chain adopted as the baseline DfT architecture. Transition delay and small delay faults are targeted using the default launch-on-capture procedure, which requires two capture cycles per pattern to launch and observe the transition at-speed; more advanced multi-capture (N-capture) sequences are not exercised. Similarly, only the ATPG tool's automatic pattern compaction is applied; more aggressive dynamic or static compaction settings are left at their default value. For small-delay fault testing, ATPG uses timing information from static timing analysis (STA) tools to execute a slack-based algorithm. With a selected time margin of 50\%, ATPG targets the top 50\% of faults with the lowest slack.

None of these capabilities are precluded by the ATPG action described in Section~\ref{sec:hammer}; rather, they represent natural directions for further experimentation on top of the benchmark suite and extended flows introduced in this paper.

\begin{table*}[!t]
\centering
\caption{Functional test programs description used for fault simulation, their observation points, average execution time, and memory footprint, grouped by functional suite.}
\label{tab:functional_programs}
\begin{tabular}{|>{\centering\arraybackslash}m{1.5cm}|>{\centering\arraybackslash}m{4.5cm}|>{\centering\arraybackslash}m{3.6cm}|>{\centering\arraybackslash}m{2.2cm}|>{\centering\arraybackslash}m{1.9cm}|>{\centering\arraybackslash}m{1.5cm}|}
\hline
\textbf{Suite} & \textbf{Configurations} & \textbf{Workload description} & \textbf{Observation points} & \begin{tabular}[c]{@{}c@{}}\textbf{Avg. Execution}\\Time \textbf{[CC]}\end{tabular} & \begin{tabular}[c]{@{}c@{}}\textbf{Memory}\\ \textbf{[KB]}\end{tabular} \\
\hline
\hline
ZEPHYR & All processor configurations & Zephyr RTOS running the standard "Hello World" application. & Processor I/Os & 116,202,623.5 & 476 \\
\hline
FFT & \texttt{FFTRocketConfig} & Fast Fourier Transform program, exercising arithmetic intensive computation and processor--accelerator data transfers. & Processor I/Os \newline Accelerator I/Os & 29,256 & 36 \\
\hline
GEMMINI & \texttt{GemminiRocketConfig} & Data alignment test program, activating complementary portions of the Gemmini datapath and its interface with the processor and memory hierarchy. & Co-processor I/Os & 16,836 & 20 \\
\hline
TEST & \texttt{AES256ECBRocketConfig} & AES-256 encryption and decryption benchmark exercising the hardware AES accelerator through repeated cryptographic operations. & Processor I/Os \newline Co-processor I/Os & 7,107,486 & 48 \\
\hline
VEC-COS & \texttt{REFV256D128M64RocketConfig} & Vectorized cosine computation benchmark exercising the RISC-V vector unit through vector floating-point arithmetic and memory operations. & Vector-Unit I/Os & 77,146,501 & 124 \\
\hline
FSWGEN & \texttt{ManyPeripheralsRocketConfig} & Device-tree-driven, automatically generated SLT exercising processor--peripheral interactions through register R/W operations, including protected register access sequences. \cite{FSWGEN} & Processor I/Os & 605,896 & 40 \\
\hline
HELLO & \begin{tabular}[c]{@{}c@{}}\texttt{DualRocketConfig}\end{tabular} & Standard Hello World benchmark with single and multi core capabilities. & Processors I/Os & 70,871 & 36 \\
\hline
\end{tabular}
\end{table*}

\subsubsection{Overall ATPG Results}
\label{sec:atpg_results}

ATPG results are reported for the 13 benchmarks of Table~\ref{tab:case_studies_architectural_description} across the three technology libraries of Table~\ref{tab:tech_libraries} and the three fault models introduced in Section~\ref{sec:hammer}, summarized in Tables~\ref{tab:atpg_nangate45}--\ref{tab:atpg_asap7}. Achieved SAF test coverage ranges from 70.43\% (\textit{IbexConfig}, ASAP7) to 98.54\% (\textit{MultiNoCConfig}, Nangate45) across the majority of case studies. TDF coverage follows a similar but consistently lower trend, from 67.04\% (\textit{IbexConfig}, ASAP7) to 98.50\% (\textit{MultiNoCConfig}, Nangate15), reflecting the additional difficulty of propagating and detecting delay defects compared to stuck-at faults. SDF coverage is substantially lower, it ranges from 29.19\% (\textit{GemminiRocketConfig}, Nangate45) to 78.68\% (\textit{MediumBoomV3Config}, Nangate15).

\begin{figure}[!htbp]
    \centering
    \includegraphics[width=1\columnwidth]{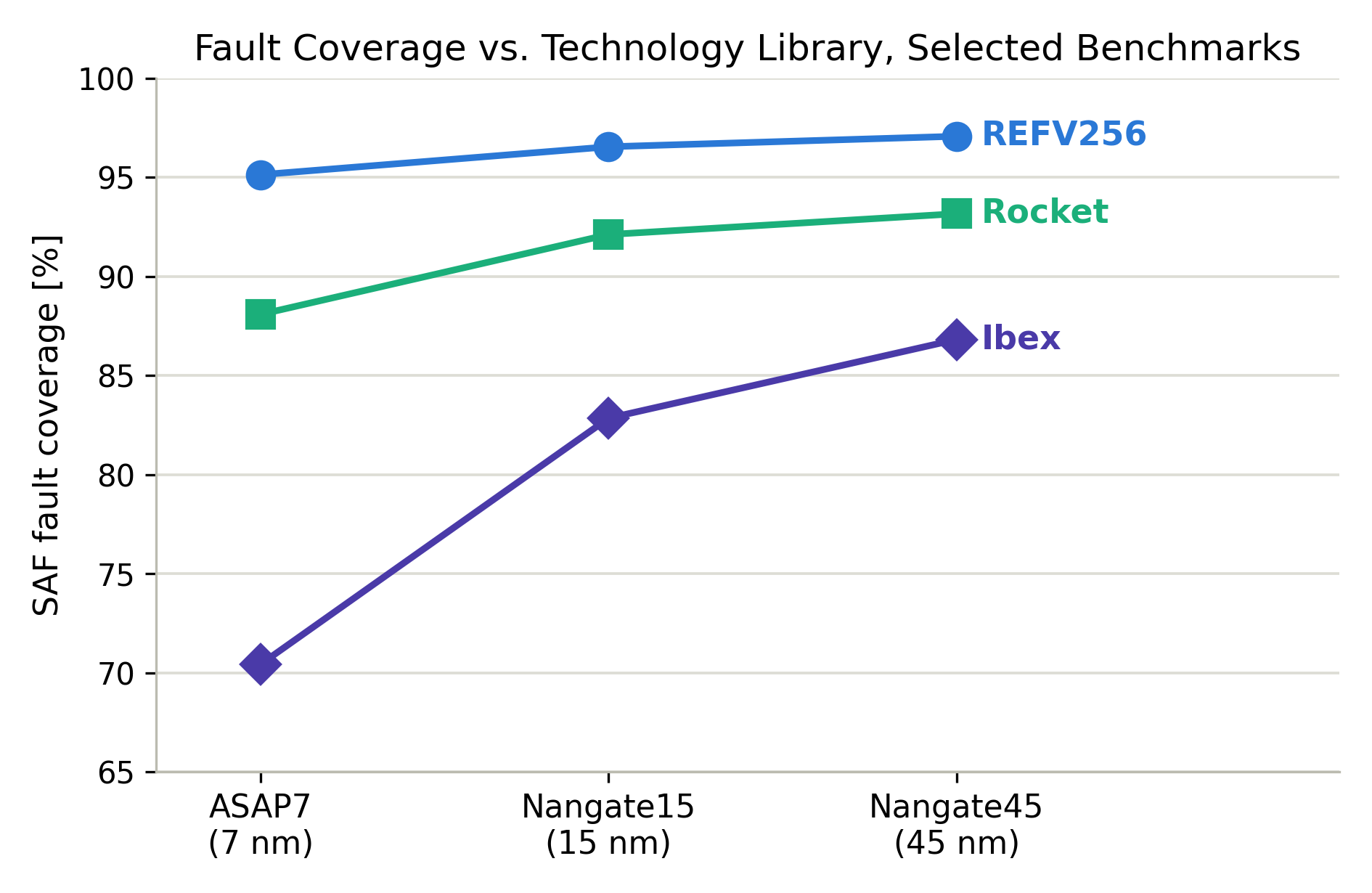}
    \caption{SAF fault coverage across technology libraries for three representative case studies: the \textit{REFV256D128M64RocketConfig} configuration, the \textit{RocketConfig} configuration, a small design with a modest penalty when using ASAP7 (\textit{IbexConfig}).}
    \label{fig:fc_tech_trend}
\end{figure}

Fault coverage also varies noticeably with the target technology library for a given design, as shown in Figure~\ref{fig:fc_tech_trend} for three representative case studies: \textit{REFV256D128M64Rocket\-Config} stays within 1.7 percentage points across all three libraries.

One configuration fall notably outside the previous mentioned ranges, that is \textit{CVA6Config} on ASAP7.

The low test coverage achieved on the CVA6 design with the ASAP7 technology library stems from both design and library constraints. To save area, ASAP7 lacks scan flip-flops with dedicated reset pins. Consequently, synthesis infers non-scan flip-flops for logic requiring asynchronous resets, rendering these core FFs untestable. Converting the reset from asynchronous to synchronous enables the ATPG tool to reach approximately 94\% coverage; however, this modification breaks logic simulation. This issue is currently under investigation jointly with the OpenHW Group.

In addition to the test coverage reported at generation time, the same patterns are re-simulated using a second commercial ATPG tool from a different vendor, with results reported in Table~\ref{tab:atpg_fault_simulation_15}; only experimental results with the Nangate15 technology library are reported for readability and for avoiding comparison between EDA tools from different vendors. This step validates the coverage figures obtained
during pattern generation and shows that the patterns produced through the ATPG action are not tied to the specific tool used to generate them, but remain usable across EDA tools from different vendors.

\subsection{Functional test evaluation}

To demonstrate the applicability of the proposed benchmark designs in \textit{Testyard} to functional fault simulation, representative software workloads are executed on each supported Chipyard configuration.

\subsubsection{Functional test programs description}

Table~\ref{tab:functional_programs} summarizes the functional test programs used for fault simulation, grouped by functional suite, along with the observation points used for good-machine/faulty-machine comparison during fault simulation campaigns. All processor configurations run the Zephyr operating system with the standard Hello World application, while specialized configurations are instead evaluated using workloads targeting their dedicated accelerators or peripherals.

\subsubsection{Overall Functional Fault Simulation Results}
\label{sec:functional_results}
The fault-simulation results are reported separately for stuck-at faults (SAF), transition-delay faults (TDF), and small-delay faults (SDF) in Table~\ref{tab:functional_all}.
For SDF testing, fault simulation campaigns are executed using a timing-annotated fault simulation. Each fault is associated with the slack value identified during the previous ATPG run, leveraging the ATPG fault list for incremental results. Each table reports, for every configuration and functional suite, the total test coverage (TC) obtained by structural tests plus the functional workload and the coverage difference with respect to structural tests (i.e., the difference between the test coverage in Table~\ref{tab:functional_all} and its corresponding entry in Table~\ref{tab:atpg_all}), denoted as $\Delta$, for the Nangate45, Nangate15, and ASAP7 technology libraries. Coverage results obtained from multiple programs belonging to the same functional suite are aggregated before being reported in Table~\ref{tab:functional_all}. Overall, these experiments demonstrate that functional fault-simulation campaigns can be performed across heterogeneous processor and accelerator Chipyard designs in feasible time.

\begin{table*}[!p]
\centering
\caption{Functional fault-simulation results for the supported configurations across the three evaluated technology libraries. For each fault model, `\#Faults' column reports the total number of simulated faults and `UT' represents the percentage of untestable faults over the corresponding fault population. `TC' denotes the functional test coverage, while `\textDelta' denotes the coverage difference with respect to ATPG.}
\label{tab:functional_all}
\setlength{\tabcolsep}{2.2pt}
\renewcommand{\arraystretch}{1.05}
\subfloat[Nangate45]{%
\resizebox{\textwidth}{!}{%
\begin{tabular}{|l|l|r|r|r|r|r|r|r|r|r|r|}
\hline
\multicolumn{1}{|c|}{\multirow{3}{*}{\textbf{Config Name}}} &
\multicolumn{1}{c|}{\multirow{3}{*}{\textbf{Suite}}} &
\multicolumn{10}{c|}{\textbf{Functional fault simulation results - Nangate45}} \\ \cline{3-12}
\multicolumn{1}{|c|}{} &
\multicolumn{1}{c|}{} &
\multicolumn{1}{c|}{\multirow{2}{*}{\textbf{\#Faults}}} &
\multicolumn{3}{c|}{\textbf{SAF}} &
\multicolumn{3}{c|}{\textbf{TDF}} &
\multicolumn{3}{c|}{\textbf{SDF}} \\ \cline{4-12}
\multicolumn{1}{|c|}{} &
\multicolumn{1}{c|}{} &
\multicolumn{1}{c|}{} &
\multicolumn{1}{c|}{\textbf{UT [\%]}} &
\multicolumn{1}{c|}{\textbf{TC [\%]}} &
\multicolumn{1}{c|}{$\bm{\Delta}$ \textbf{[\%]}} &
\multicolumn{1}{c|}{\textbf{UT [\%]}} &
\multicolumn{1}{c|}{\textbf{TC [\%]}} &
\multicolumn{1}{c|}{$\bm{\Delta}$ \textbf{[\%]}} &
\multicolumn{1}{c|}{\textbf{UT [\%]}} &
\multicolumn{1}{c|}{\textbf{TC [\%]}} &
\multicolumn{1}{c|}{$\bm{\Delta}$ \textbf{[\%]}} \\ \hline \hline
AES256ECBRocketConfig & TEST & 7,354,394 & 0.92 & 98.42 & 2.82 & 0.94 & 97.90 & 2.68 & 0.69 & 72.74 & 5.53 \\
\cline{1-1}\cline{3-12}
MediumBoomV3Config & \multirow{2}{*}{ZEPHYR}  & 6,558,356 & 1.07 & 97.93 & 1.30 & 1.09 & 96.30 & 1.55 & 0.30 & 72.83 & 6.27 \\
\cline{1-1}\cline{3-12}
MediumBoomV4Config &  & 6,957,208 & 1.47 & 98.20 & 0.90 & 1.54 & 94.80 & 1.18 & 0.47 & 69.09 & 5.69 \\
\hline
GemminiRocketConfig & GEMMINI & 9,955,836 & 2.02 & 94.20 & 0.60 & 2.18 & 92.70 & 0.35 & 1.83 & 51.67 & 0.52 \\
\hline
IbexConfig & ZEPHYR & 1,049,732 & 1.79 & 92.18 & 5.36 & 1.85 & 89.03 & 4.80 & 0.13 & 73.11 & 7.74 \\
\hline
FFTRocketConfig & FFT & 2,158,410 & 1.13 & 95.45 & 2.06 & 1.15 & 93.77 & 1.83 & 0.58 & 68.31 & 3.74 \\
\hline
ShuttleConfig & \multirow{3}{*}{ZEPHYR} & 2,948,170 & 1.41 & 96.16 & 2.29 & 1.44 & 94.31 & 1.94 & 0.38 & 70.34 & 4.59 \\
\cline{1-1}\cline{3-12}
MultiNoCConfig &  & 36,822,786 & 0.97 & 98.91 & 0.37 & 0.98 & 98.82 & 0.32 & 0.21 & 65.12 & 1.40 \\
\cline{1-1}\cline{3-12}
RocketConfig &  & 2,108,074 & 1.08 & 95.32 & 2.15 & 1.11 & 93.70 & 1.95 & 0.58 & 68.92 & 4.10 \\
\hline
ManyPeripheralsRocketConfig & FSWGEN & 2,653,752 & 1.02 & 96.37 & 1.77 & 1.04 & 94.86 & 1.63 & 0.53 & 69.96 & 3.48 \\
\hline
REFV256D128M64RocketConfig & VEC-COS & 5,391,968 & 2.19 & 98.09 & 1.01 & 2.23 & 97.24 & 0.95 & 1.87 & 68.93 & 10.74 \\
\hline
DualRocketConfig & mt-hello & 3,225,532 & 1.05 & 96.71 & 1.76 & 1.07 & 95.44 & 1.61 & 0.66 & 68.29 & 4.44 \\
\hline
\end{tabular}%
}
\label{tab:functional_nangate45_results}
}
\vspace{1em}
\subfloat[Nangate15]{%
\resizebox{\textwidth}{!}{%
\begin{tabular}{|l|l|r|r|r|r|r|r|r|r|r|r|}
\hline
\multicolumn{1}{|c|}{\multirow{3}{*}{\textbf{Config Name}}} &
\multicolumn{1}{c|}{\multirow{3}{*}{\textbf{Suite}}} &
\multicolumn{10}{c|}{\textbf{Functional fault simulation results - Nangate15}} \\ \cline{3-12}
\multicolumn{1}{|c|}{} &
\multicolumn{1}{c|}{} &
\multicolumn{1}{c|}{\multirow{2}{*}{\textbf{\#Faults}}} &
\multicolumn{3}{c|}{\textbf{SAF}} &
\multicolumn{3}{c|}{\textbf{TDF}} &
\multicolumn{3}{c|}{\textbf{SDF}} \\ \cline{4-12}
\multicolumn{1}{|c|}{} &
\multicolumn{1}{c|}{} &
\multicolumn{1}{c|}{} &
\multicolumn{1}{c|}{\textbf{UT [\%]}} &
\multicolumn{1}{c|}{\textbf{TC [\%]}} &
\multicolumn{1}{c|}{$\bm{\Delta}$ \textbf{[\%]}} &
\multicolumn{1}{c|}{\textbf{UT [\%]}} &
\multicolumn{1}{c|}{\textbf{TC [\%]}} &
\multicolumn{1}{c|}{$\bm{\Delta}$ \textbf{[\%]}} &
\multicolumn{1}{c|}{\textbf{UT [\%]}} &
\multicolumn{1}{c|}{\textbf{TC [\%]}} &
\multicolumn{1}{c|}{$\bm{\Delta}$ \textbf{[\%]}} \\ \hline \hline
AES256ECBRocketConfig & TEST & 7,702,008 & 2.86 & 98.20 & 2.83 & 4.90 & 97.53 & 2.74 & 0.69 & 74.21 & 4.96 \\
\cline{1-1}\cline{3-12}
MediumBoomV3Config & \multirow{2}{*}{ZEPHYR} & 7,022,048 & 2.97 & 97.60 & 1.20 & 4.99 & 96.06 & 1.32 & 0.27 & 73.79 & 4.83 \\
\cline{1-1}\cline{3-12}
MediumBoomV4Config &  & 7,566,686 & 3.41 & 96.76 & 1.13 & 5.54 & 94.28 & 1.22 & 0.47 & 70.07 & 5.12 \\
\hline
GemminiRocketConfig & GEMMINI & 10,681,356 & 3.51 & 94.10 & 0.54 & 5.13 & 93.08 & 0.30 & 1.91 & 53.14 & 0.50 \\
\hline
IbexConfig & ZEPHYR & 1,149,276 & 4.06 & 88.58 & 5.73 & 6.42 & 85.15 & 4.27 & 0.12 & 72.79 & 6.70 \\
\hline
FFTRocketConfig & FFT & 2,294,346 & 2.93 & 94.22 & 1.93 & 4.86 & 92.80 & 1.75 & 0.57 & 69.04 & 3.50 \\
\hline
ShuttleConfig & \multirow{3}{*}{ZEPHYR} & 3,105,074 & 3.30 & 95.24 & 2.12 & 5.31 & 93.57 & 1.75 & 0.38 & 70.24 & 4.56 \\
\cline{1-1}\cline{3-12}
MultiNoCConfig &  & 38,532,222 & 2.52 & 98.89 & 0.36 & 4.08 & 98.80 & 0.30 & 0.24 & 61.72 & 1.41 \\
\cline{1-1}\cline{3-12}
RocketConfig &  & 2,242,062 & 2.89 & 94.20 & 2.08 & 4.83 & 92.76 & 1.87 & 0.57 & 69.70 & 3.69 \\
\hline
ManyPeripheralsRocketConfig & FSWGEN & 2,792,874 & 2.99 & 95.45 & 1.75 & 5.06 & 94.05 & 1.62 & 0.53 & 70.18 & 3.27 \\
\hline
REFV256D128M64RocketConfig & VEC-COS & 5,617,026 & 3.46 & 97.54 & 0.99 & 4.83 & 96.79 & 0.94 & 1.87 & 69.36 & 9.29 \\
\hline
DualRocketConfig & mt-hello & 3,429,820 & 2.75 & 95.94 & 1.68 & 4.55 & 94.78 & 1.56 & 0.63 & 67.64 & 4.34 \\
\hline
\end{tabular}%
}
\label{tab:functional_nangate15_results}
}
\vspace{1em}
\subfloat[ASAP7]{%
\resizebox{\textwidth}{!}{%
\begin{tabular}{|l|l|r|r|r|r|r|r|r|r|r|r|}
\hline
\multicolumn{1}{|c|}{\multirow{3}{*}{\textbf{Config Name}}} &
\multicolumn{1}{c|}{\multirow{3}{*}{\textbf{Suite}}} &
\multicolumn{10}{c|}{\textbf{Functional fault simulation results - ASAP7}} \\ \cline{3-12}
\multicolumn{1}{|c|}{} &
\multicolumn{1}{c|}{} &
\multicolumn{1}{c|}{\multirow{2}{*}{\textbf{\#Faults}}} &
\multicolumn{3}{c|}{\textbf{SAF}} &
\multicolumn{3}{c|}{\textbf{TDF}} &
\multicolumn{3}{c|}{\textbf{SDF}} \\ \cline{4-12}
\multicolumn{1}{|c|}{} &
\multicolumn{1}{c|}{} &
\multicolumn{1}{c|}{} &
\multicolumn{1}{c|}{\textbf{UT [\%]}} &
\multicolumn{1}{c|}{\textbf{TC [\%]}} &
\multicolumn{1}{c|}{$\bm{\Delta}$ \textbf{[\%]}} &
\multicolumn{1}{c|}{\textbf{UT [\%]}} &
\multicolumn{1}{c|}{\textbf{TC [\%]}} &
\multicolumn{1}{c|}{$\bm{\Delta}$ \textbf{[\%]}} &
\multicolumn{1}{c|}{\textbf{UT [\%]}} &
\multicolumn{1}{c|}{\textbf{TC [\%]}} &
\multicolumn{1}{c|}{$\bm{\Delta}$ \textbf{[\%]}} \\ \hline \hline
AES256ECBRocketConfig & TEST & 5,467,150 & 1.42 & 96.91 & 3.93 & 1.52 & 91.60 & 1.01 & 0.90 & 59.35 & 0.04 \\
\cline{1-1}\cline{3-12}
MediumBoomV3Config & \multirow{2}{*}{ZEPHYR} & 5,482,378 & 1.46 & 94.92 & 4.04 & 1.60 & 87.37 & 1.63 & 0.30 & 53.60 & 0.04 \\
\cline{1-1}\cline{3-12}
MediumBoomV4Config &  & 5,851,724 & 2.00 & 96.31 & 3.03 & 2.15 & 89.32 & 1.19 & 0.18 & 56.21 & 0.06 \\
\hline
GemminiRocketConfig & GEMMINI & 9,047,954 & 2.91 & 91.43 & 1.04 & 3.15 & 89.03 & 0.54 & 3.22 & 43.35 & 0.00 \\
\hline
IbexConfig & ZEPHYR & 759,366 & 2.94 & 82.48 & 12.05 & 3.50 & 75.66 & 8.62 & 0.23 & 46.87 & 0.36 \\
\hline
FFTRocketConfig & FFT & 1,800,798 & 1.92 & 92.45 & 5.86 & 2.16 & 85.98 & 3.20 & 0.33 & 49.80 & 0.01 \\
\hline
ShuttleConfig & \multirow{3}{*}{ZEPHYR} & 2,360,872 & 2.04 & 93.87 & 3.86 & 2.24 & 87.73 & 1.31 & 0.12 & 54.42 & 0.00 \\
\cline{1-1}\cline{3-12}
MultiNoCConfig &  & 30,425,280 & 1.33 & 97.70 & 1.24 & 1.38 & 95.51 & 0.71 & 0.25 & 72.50 & 0.00 \\
\cline{1-1}\cline{3-12}
RocketConfig &  & 1,720,516 & 1.90 & 92.50 & 4.41 & 2.14 & 85.83 & 1.55 & 0.16 & 52.71 & 0.00 \\
\hline
ManyPeripheralsRocketConfig & FSWGEN & 2,623,452 & 1.12 & 93.49 & 2.92 & 1.18 & 88.45 & 0.96 & 0.38 & 57.83 & 0.02 \\
\hline
REFV256D128M64RocketConfig & VEC-COS & 4,771,272 & 2.38 & 97.07 & 1.93 & 2.54 & 92.42 & 0.72 & 1.31 & 56.40 & 0.01 \\
\hline
DualRocketConfig & mt-hello & 2,675,988 & 1.64 & 93.91 & 3.54 & 1.83 & 87.78 & 1.32 & 0.17 & 52.01 & 0.00 \\
\hline
\end{tabular}%
}
\label{tab:functional_asap7_results}
}
\end{table*}

\begin{table*}[!t]
\centering
\caption{Average functional fault simulation campaign execution times in minutes across the three evaluated technology libraries for each hardware configuration and workload suite.}
\label{tab:avg_fsim_execution_times}
\begin{tabular}{|l|c|c|c|c|}
\hline
\textbf{Config Name} & \textbf{Suite} & \textbf{Nangate45 [min]} & \textbf{Nangate15 [min]} & \textbf{ASAP7 [min]} \\
\hline
\hline
AES256ECBRocketConfig & TEST & 1970 & 2200 & 7080 \\
\hline
CVA6Config & \multirow{3}{*}{ZEPHYR} & 2990 & 3360 & 12,530 \\
\cline{1-1}\cline{3-5}
MediumBoomV3Config & & 1,280 & 1,580 & 1,180 \\
\cline{1-1}\cline{3-5}
MediumBoomV4Config &  & 1,590 & 2,170 & 1,310 \\
\hline
GemminiRocketConfig & GEMMINI & 670 & 450 & 200 \\
\hline
IbexConfig & ZEPHYR & 270 & 280 & 1,060 \\
\hline
FFTRocketConfig & FFT & 10 & 10 & 20 \\
\hline
ShuttleConfig & \multirow{3}{*}{ZEPHYR} & 400 & 520 & 280 \\
\cline{1-1}\cline{3-5}
MultiNoCConfig &  & 3,650 & 3,580 & 37,440 \\
\cline{1-1}\cline{3-5}
RocketConfig &  & 180 & 190 & 240 \\
\hline
ManyPeripheralsRocketConfig & FSWGEN & 40 & 50 & 690 \\
\hline
REFV256D128M64RocketConfig & COS-VEC & 880 & 1,070 & 1310 \\
\hline
DualRocketConfig & HELLO & 40 & 60 & 50 \\
\hline
\end{tabular}
\end{table*}

\subsubsection{Overall Functional Fault Simulation Execution Time}

In order to assess the feasibility of fault simulations on big designs and long functional test programs, Table~\ref{tab:avg_fsim_execution_times} reports the average execution time (in minutes) required to complete a functional fault simulation campaign across the three evaluated technology libraries: Nangate45, Nangate15, and ASAP7. For each target processor configuration and its corresponding functional test suite (e.g., Zephyr, Gemmini, FFT), the table details the compute overhead associated with running the simulation campaigns. Multi-row entries group configurations that share the same underlying workload suite to facilitate a direct comparison of runtime scaling relative to design complexity and technology library node.

\section{Conclusions}
\label{sec:conclusions}

This work addressed two critical needs in the hardware testing research community: the availability of modern, open-source benchmarks of industrial complexity, and a unified agile framework for prototyping structural and functional testing methodologies.

The Hammer framework was extended with DfT insertion, ATPG, and functional fault simulation actions, each configurable through high-level YAML descriptions that abstract the underlying EDA-tool-specific commands. By reducing the engineering effort required to adapt fragile, tool-dependent scripts, the extended framework allows researchers to focus on methodology rather than infrastructure, significantly accelerating the prototyping of new hardware testing approaches.

The extended framework was applied to a representative subset of SoC configurations available in the Chipyard ecosystem, spanning from simple microcontroller configurations to complex out-of-order and accelerator-equipped designs. Structural and functional test results were reported for stuck-at, transition delay, and small delay fault models across three technology nodes: the Nangate 45$nm$ Planar node, the Nangate 15$nm$ FinFET node~\cite{SILVACO_OPENPDK}, and the advanced ASAP7 7$nm$ FinFET node~\cite{ASAP7}.

The results confirm that the generated Testyard benchmarks represent a significant step forward in complexity with respect to traditional academic suites such as ITC-99, offering a fault universe several orders of magnitude larger and a design structure that more faithfully reflects the challenges of modern industrial SoCs. This is important because some recent works have found that the effectiveness of some testing methodologies can vary between smaller benchmarks and more complex industrial designs~\cite{BERNARDI_LOGICDIAGNOSIS_MDPI_2024, IARIA_COSTMODEL_TOC_2025}. For example, logic diagnosis on small designs can be much easier due to the reduced aliasing in the output signature~\cite{BERNARDI_LOGICDIAGNOSIS_MDPI_2024, BERNARDI_ENHANCINGLOGICDIAGNOSIS_LATS_2025}. This highlights the value of validating methodologies on benchmarks of varying complexity. Moreover, if a method relies on a uniform distribution of defect probability, it can be less effective on large designs, where gate density plays a key role~\cite{BERNARDI_LAYOUTCHARACTERISTICS_IOLTS_2023, IARIA_COSTMODEL_TOC_2025}.

Industrial designs vary significantly in complexity depending on the target application, typically ranging from 2 million gates~\cite{CIULLO_WEAKDETECTIONS_ITCASIA_2025} to 20 million gates in high-performance automotive SoCs~\cite{ANGIONE_SLTXBAR_TOC_2024, ANGIONE_SLTCAN_TOC_2025, FILIPPONI_AMU_TOC_2026}, and reaching upwards of 90 million gates for large-scale SoCs in Advanced Driver Assistance Systems (ADAS). The flexibility of the \textit{Testyard} framework allows researchers to select configurations that mimic these industrial use cases by adjusting core counts and integrated modules. While circuits like \textit{c7552} from ISCAS-85~\cite{ISCAS85} or \textit{b18} from ITC-99~\cite{ITC99} have served the community for decades, they no longer represent the gate-level density or memory complexity of contemporary silicon: the largest case study evaluated in this work, \textit{MultiNoCConfig}~\cite{CHIPYARD}, already reaches a fault universe 172 times larger than \textit{b18}. By integrating additional cores and peripherals, these designs can be expanded further to effectively close the complexity gap with the largest industry-standard devices.

To foster reproducibility and further adoption by the research community of the \textit{Testyard} framework, the synthesized DfT-inserted netlists, the extended
Hammer flows, and the scripts are publicly released on
GitHub~\cite{CHIPYARD_CAD, HAMMER_CAD}.

\section*{Acknowledgment}
The authors used OpenAI’s ChatGPT Edu to improve the grammar and clarity of the article, all technical content was authored and verified by the authors.

\bibliographystyle{IEEEtran}
\bibliography{biblio}

\begin{IEEEbiography}[{\includegraphics[width=1in,height=1.25in,clip,keepaspectratio]{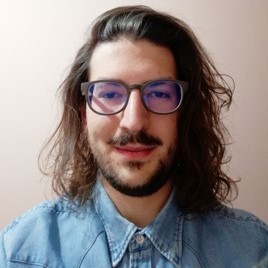}}]{FRANCESCO ANGIONE} 
(Member, IEEE) is a Computer Engineer with an M.Sc. in Embedded Systems track, obtained from Politecnico di Torino in 2020. He holds a Ph.D. on "System-Level-Test techniques for Automotive SoCs" at Politecnico di Torino in the CAD \& Reliability group. His main interests are real-time operating systems, computer architectures, and their dependability.
\end{IEEEbiography}

\begin{IEEEbiography}[{\includegraphics[width=1in,height=1.25in,clip,keepaspectratio]{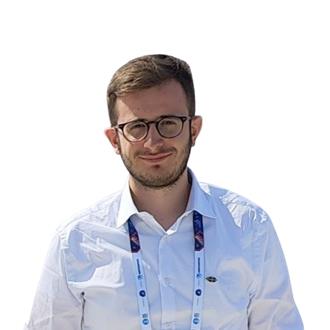}}]{NICOLA DI GRUTTOLA GIARDINO} (Graduate Student Member, IEEE) received the M.Sc. degree in computer engineering, embedded systems track from Politecnico di Torino, in 2024. He is currently working toward the Ph.D. degree in aerospace engineering. His main interests are digital design, real-time operating systems, and firmware development. He is part of the Electronic CAD \& Reliability Group, as well as the STAR Team. 
\end{IEEEbiography}

\begin{IEEEbiography}[{\includegraphics[width=1in,height=1.25in,clip,keepaspectratio]{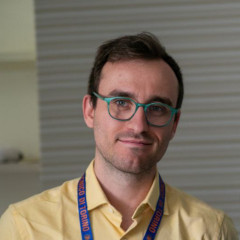}}]{GABRIELE FILIPPONI} 
(Member, IEEE) is a third-year Ph.D. student in Computer and Control Engineering at Politecnico di Torino. He is part of the Electronic CAD \& Reliability group, with a focus on Embedded electronics testing. His main field of research regards ultra reliability for in-field Automotive SoC, specifically exploiting the LBIST.
\end{IEEEbiography}

\begin{IEEEbiography}[{\includegraphics[width=1in,height=1.25in,clip,keepaspectratio]{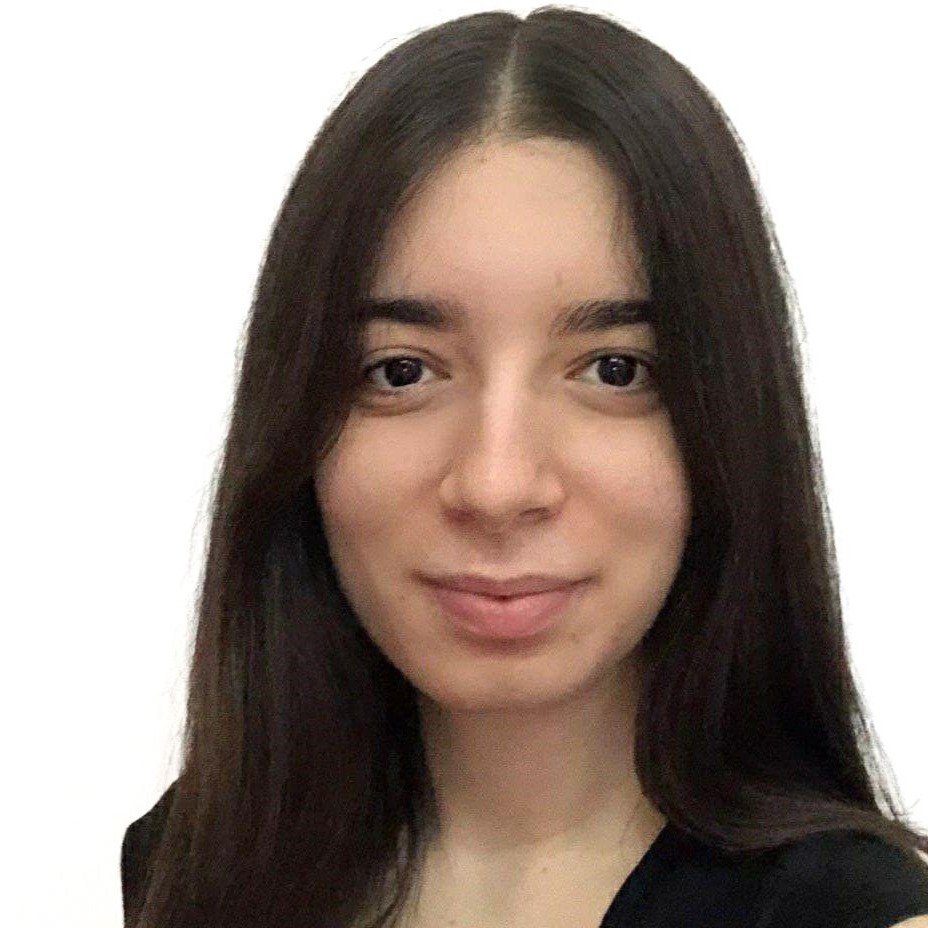}}]{GIUSY IARIA} 
(Member, IEEE) is a Research Fellow at Politecnico di Torino, Italy, where she obtained her PhD in 2025 in Computer and Control Engineering. She is part of the CAD \& Reliability research group with a strong focus on Embedded electronics testing for large Automotive devices.
\end{IEEEbiography}

\begin{IEEEbiography}[{\includegraphics[width=1in,height=1.25in,clip,keepaspectratio]{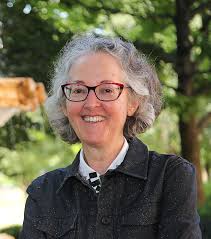}}]{IRITH POMERANZ} 
(Fellow, IEEE) received
the B.Sc. degree (summa cum laude) in computer engineering and the D.Sc. degree from the Department of Electrical Engineering, Technion-Israel Institute of Technology, in 1985 and 1989,
respectively.
From 1989 to 1990, she was a Lecturer with
the Department of Computer Science, Technion.
From 1990 to 2000, she was a Faculty Member
with the Department of Electrical and Computer
Engineering, The University of Iowa. In 2000, she joined Purdue University,
West Lafayette, IN, USA, where she is the Cadence Professor in the Elmore
Family School of Electrical and Computer Engineering.
Prof. Pomeranz is a Golden Core Member of the IEEE Computer Society.
She was a recipient of the NSF Young Investigator Award, in 1993,
and The University of Iowa Faculty Scholar Award, in 1997. She was
one of the very first three featured authors on IEEE Xplore, posted in
February 2020, and was named a ScholarGPS Highly Ranked Scholar,
in 2024. She was awarded the IEEE Computer Society Test Technology
Technical Council (TTTC) Lifetime Contribution Medal, in 2025. Three
of her conference papers won best paper awards, and four other papers
were nominated for best paper awards. One of the papers she co-authored
was selected by the 2016 International Test Conference as the most
significant paper published ten years before. She delivered a keynote
speech at the 2006 Asian Test Symposium, and the 2025 Data Integrity
and Secure Cloud Computing Workshop. She served as a Guest Editor
for IEEE TRANSACTIONS ON COMPUTERS, in January 1998, special issue on
Dependability of Computing Systems, and as the Program Co-Chair of
the 1999 Fault-Tolerant Computing Symposium. She served as the Program
Chair of the 2004 and 2005 VLSI Test Symposium, and the General Chair of
the 2006 VLSI Test Symposium. She served as an Associate Editor for ACM
Transactions on Design Automation, IEEE Transactions on Computers, and IEEE Transactions on Very Large Scale Integration Systems.
\end{IEEEbiography}

\begin{IEEEbiography}[{\includegraphics[width=1in,height=1.25in,clip,keepaspectratio]{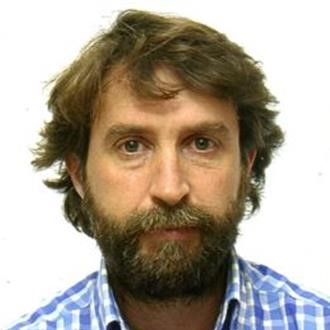}}]{PAOLO BERNARDI} 
(Senior Member, IEEE) (MS'02 and PhD'06 in Computer Science) is an Associate Professor of Politecnico di Torino University, where he works in the Electronic CAD and Reliability research group. His current interests include System-on-Chip tests and reliability. Prof. Bernardi is the Program Chair of the IEEE International Test Conference 2025 and he has been the General Chair of the IEEE European Test Symposium 2023. 
\end{IEEEbiography}

\end{document}